\documentclass[usenatbib]{mn2e}
\usepackage{times}
\usepackage{epsfig}
\usepackage{amsmath}
\title[Transient galaxies in groups at $z=1$]{Direct observational
  evidence for a large transient galaxy
  population in groups at $0.85<z<1$}
\author[Balogh et al.]{Michael L. Balogh$^{1}$, Sean L. McGee$^{1,2}$,
  David J. Wilman$^{3}$, Alexis Finoguenov$^{3,4}$,
\newauthor Laura C. Parker$^{5}$, Jennifer L. Connelly$^{3}$,
John~S. Mulchaey$^{6}$, Richard G. Bower$^{2}$, 
\newauthor Masayuki Tanaka$^{7}$, Stefania Giodini$^{8}$ 
\\
$^{1}$Department of Physics and Astronomy, University of Waterloo, Waterloo, Ontario, N2L 3G1, Canada\\
$^{2}$Department of Physics, University of Durham, Durham, UK, DH1 3LE\\
$^{3}$Max--Planck--Institut f{\" u}r extraterrestrische Physik, Giessenbachstrasse 85748 Garching Germany\\
$^{4}$CSST, University of Maryland, Baltimore County, 1000 Hilltop Circle,  Baltimore, MD 21250, USA\\
$^{5}$Department of Physics and Astronomy, McMaster University, Hamilton, Ontario, L8S 4M1 Canada\\
$^{6}$Observatories of the Carnegie Institution, 813 Santa Barbara Street, Pasadena, California, USA\\
$^{7}$Institute for the Physics and Mathematics of the Universe, University of Tokyo, Kashiwa 2778582, Japan\\
$^{8}$Leiden Observatory, Leiden University, PO Box 9513, 2300 RA Leiden,
the Netherlands\\
}
\date{\today}

\def\gtrsim{\mathrel{\raise0.35ex\hbox{$\scriptstyle >$}\kern-0.6em
\lower0.40ex\hbox{{$\scriptstyle \sim$}}}}
\def\lesssim{\mathrel{\raise0.35ex\hbox{$\scriptstyle <$}\kern-0.6em
\lower0.40ex\hbox{{$\scriptstyle \sim$}}}}

\begin{document} 
\maketitle

\begin{abstract}
We introduce our survey of galaxy groups at $0.85<z<1$, as an extension of the Group Environment and
Evolution Collaboration (GEEC).   Here we present the first results, based on
Gemini GMOS-S nod-and-shuffle spectroscopy of seven galaxy groups selected from spectroscopically
confirmed, extended {\it XMM} detections in COSMOS.  We use photometric redshifts
to select potential group members for spectroscopy, and target galaxies
with $r<24.75$.  In total we have over 100 confirmed group members, and
four of the groups have $>15$ members.  The dynamical mass estimates
are in good agreement with the masses estimated from the X--ray
luminosity, with most of the groups having $13<\log{M_{\rm
    dyn}/M_\odot}<14$.   We compute stellar masses by template-fitting
the spectral energy distributions; our spectroscopic sample is
statistically complete for all galaxies with $M_{\rm
  star}\gtrsim10^{10.1}M_\odot$, and for blue galaxies we sample masses as
low as $M_{\rm star}\sim10^{8.8}M_\odot$.  The fraction of total mass in
galaxy starlight spans a range of 0.25--3 per cent, for the six groups with reliable mass
determinations.  Like lower-redshift groups, these systems are
dominated by red galaxies, at all stellar masses $M_{\rm
  star}>10^{10.1}M_\odot$.   Few group galaxies inhabit the ``blue cloud''
that dominates the surrounding field; instead, we find a 
large and possibly distinct population of galaxies with intermediate colours.  The ``green
valley'' that exists at low redshift is instead well-populated in these
groups, containing $\sim 30$ per cent of the galaxies.   These do not
appear to be exceptionally dusty galaxies, and about half show prominent
Balmer-absorption lines.  
Furthermore, their {\it HST} morphologies appear to be
intermediate between those of red-sequence and blue-cloud galaxies of
the same stellar mass.  Unlike red-sequence galaxies, most of the green
galaxies have a disk component, but one that is
smaller and less structured than disks found in the blue cloud.  We
postulate that these are a transient population, migrating from the
blue cloud to the red sequence, 
with a star formation rate that declines with an exponential timescale $0.6\mbox{Gyr}<\tau<2\mbox{Gyr}$.
Such galaxies may not be exclusive to the group environment, as we find
examples also amongst the non-members.  However, their 
prominence
among the group galaxy population, and the marked lack of blue,
star-forming galaxies, provides evidence
that the group environment either directly reduces star formation in
member galaxies, or
at least prevents its rejuvenation during the normal cycle of galaxy evolution.
\end{abstract}

\section{Introduction}\label{sec-intro}
Galaxy evolution since about $z\sim 2$ is characterized generally by
the cessation of star formation.  On average, the star formation rate
density in the Universe has declined by a factor $\sim 10$ since that
time \citep[e.g.][]{L96,Hopkins04,ROLES_Gilbank}.  The rate of decline
appears to depend on both stellar mass
\citep[e.g.][]{Cowie+96,Juneau+04,Noeske07,Bell+07,ROLES_Gilbank} and
environment
\citep[e.g.][]{Pogg05,DEEPII_envt_again,zCOSMOS_Maier,zCOSMOS_Iovino,V+10,zCOSMOS_Peng,zCOSMOS_Cucciati,Sobral,GEEC_sedfit}.
This leads to a fairly consistent empirical description, at least for
massive galaxies.  The galaxy population is well-modeled by
having star formation ``quenched'' on relatively short timescales; this
quenching happens first for the most massive galaxies, and later for
lower--mass systems \citep[e.g.][]{Noeske2}.  This evolution is
somewhat accelerated in denser environments
\citep{zCOSMOS_Iovino,V+10,ROLES_tornado}, especially for the
lowest-mass galaxies studied to date.

However, we lack a clear understanding of the physical processes
driving this evolution.  It appears that the dependence on
stellar--mass and environment are separable \citep{BaldryV}, and likely
point to different mechanisms.  Recently, \citet{zCOSMOS_Peng} have
proposed an interesting empirical description of this, where the
quenching rate is described by one term that is proportional to star
formation rate, and another that is related to the local density.  This
works surprisingly well, but still doesn't lend itself to a clear
physical interpretation.

Galaxy formation models are too complex and under-constrained to provide
an unequivocal answer at this time.  However, they do indicate that
parametric treatments of energy input due to supernova and
supermassive black hole accretion can do a reasonable job of explaining
the mass--dependent quenching \citep[e.g.][]{Croton05,Bower+08}.  The
environmental term is accounted for to some extent in the so--called
``halo model'', where the central galaxy of a given dark matter halo is
treated somehow differently from the satellites
\citep[e.g.][]{GB08,SS09,McGee-accretion}.   Such a distinction is expected
at some level, since the hot gas that is thought to fill a
common halo should, for the most part, cool and condense only onto the
central galaxy.  Much of the hot gas associated with satellite galaxies will
likely get tidally stripped, ram-pressure stripped, or shock heated
until it becomes part of the common halo \citep[e.g.][]{Ian_rps,KM07,Font+08}.  This picture relies on a
number of untested assumptions, however, and simple implementations of
this do not do a very good job of predicting the observed star
formation histories of grouped galaxies
\citep{Weinmann_1+06,GB08,geec_colours}.   

The study of galaxy evolution in dense environments has been hampered
by several observational difficulties.  The easiest place to look is in
the cores of rich clusters at relatively low redshift \citep[e.g.][and
many others]{B+97,P+99,A1689,2dF,Sloan_sfr_short,LGBB}.  But the cores of
rich clusters are extreme environments where, for example, ram-pressure
stripping can remove even the cold gas from disks \citep[e.g.][]{QMB,KK04},
and galaxy interactions effectively distort galaxy morphologies
\citep{Moore+96}.  These are interesting processes, but are unlikely to
be the explanation for trends observed in lower--density environments
\citep[e.g.][]{2dfsdss}.  Moreover, low redshift systems provide
additional challenges.  Galaxies at low redshift are generally older and more gas poor,
due to normal evolution, so there is less potential for environment to
disrupt star formation.  Also, the accretion rate
of new, relatively gas--rich galaxies, into clusters is lower than
it was in the past \citep[e.g.][]{Erica,McGee-accretion}, and
recently accreted galaxies will represent a small fraction of the
entire population which has built up over a Hubble time.  So there is
less opportunity to catch galaxies in the act of transformation, if
such a process occurs.

Thus, the best place to observe environmentally--driven evolution in
action is in gas-rich, low--mass galaxies inhabiting small groups, at
relatively high redshift.  This was the motivation behind the Group
Environment and Evolution Collaboration (GEEC), that began with deep
Magellan LDSS3 spectroscopy of groups at $0.3<z<0.55$, selected from the CNOC2 survey
fields \citep{CNOC2_groupsI,CNOC2_groupsII}.  In a series of papers we
showed that while disk--dominated, star forming galaxies of a given
stellar mass are less
common in groups than in the general field,  the properties of the star--forming
population itself seems independent of environment \citep[e.g.][]{GEEC_sedfit}.  This is
surprising, as it suggests that either the transformation is very
rapid, or the number of galaxies undergoing transformation at the epoch
of observation is small.  At the same time, group galaxies look much
more like the general field in most respects than is the case at lower
redshift \citep{GEEC_sedfit}.  This suggests that evolution in groups
is a relatively recent phenomenon \citep[see also ][]{LGBB}.

Nonetheless, groups at $z\sim 1$ and beyond still show a larger
population of passively--evolving galaxies than the surrounding field
\citep[e.g.][]{Cooper+10}.  Interestingly, however, there is increasing evidence
that those galaxies that {\it are} forming stars, are doing so at a higher
rate than field galaxies of similar stellar mass \citep{Elbaz+07,Muzzin+08,DEEPII_envt_again,I+09,Sobral,ROLES_tornado,K+10}.   It may be
at this epoch that dense environments stimulate star formation,
leading to a rapid consumption of gas and leading to the dominant old
population at $z=0$.   The joint density- and mass-dependence of this
effect is nicely illustrated by \citet{Sobral}.  The higher accretion rate, and younger age of
the Universe, has led us to predict that there will be more diversity
among galaxy groups at this redshift \citep{McGee-accretion}.  

We have thus embarked on an ambitious spectroscopic campaign, using the
GMOS spectrographs on Gemini North and South to study the galaxy
populations in $\sim 20$ groups, mostly selected within the COSMOS
\citep{COSMOS} survey fields.  The deep spectroscopy
achieves two main goals.  First, we can study galaxies with low stellar
mass; our sample is complete for {\it red} galaxies 1.0 mag fainter
than the zCOSMOS 10k survey \citep{zCOSMOS}, 0.65 mag fainter than
DEEP2 \citep{DEEPII_envt_again} and $\sim 0.25$ mag fainter
than EDisCS at this redshift \citep{Halliday+04}; for blue galaxies we
are complete to even fainter magnitudes.  This depth is vital for studying
environmental effects, which appear to be most important for the
lowest--mass galaxies \citep{zCOSMOS_Cucciati}.
Secondly, the depth allows us to obtain $\sim 20$ spectroscopic members per
group, which is sufficient to estimate the dynamical mass, and
dynamical state, of each group \citep{Hou09}.  Moreover, it allows us to consider the
galaxy population of individual groups, or subsets of groups.  Thus, we
can attempt to link transient galaxies to the dynamics or recent
accretion history of the larger--scale environment.

We present our data, including a full description of the group and
spectroscopic selection, in \S~\ref{sec-data}.  The basic analysis of
redshift and stellar mass measurements, and determination of group
dynamical masses, is presented in \S~\ref{sec-anal}.  Our main results,
on the stellar fraction and the discovery of a dominant group
population of galaxies with intermediate colours, spectral type and
morphology, are presented in \S~\ref{sec-results}.  Finally, we discuss
the implications of our results, and summarize our conclusions, in
\S~\ref{sec-discuss}.  

Throughout the paper, we assume a cosmology with $\Omega_m=0.3$,
$\Omega_\Lambda=0.7$ and $h=H_\circ/\left(100
  \mbox{km}/\mbox{s}/\mbox{Mpc}\right)=0.7$.  All magnitudes are on the
AB system.

\section{Data}\label{sec-data}
\subsection{The parent catalogues}
Our survey targets galaxy groups within the COSMOS field
\citep{COSMOS}.  This field benefits from a wide range of publicly
available, deep, multiwavelength photometry.  The backbone is the
widest {\it Hubble Space Telescope} (HST) survey ever undertaken, covering
$\sim 2$ square degrees with the F814W filter on the {\it Advanced
  Camera for Surveys} \citep{COSMOS_hst}.   In addition, we will make
use of the
deep X-ray observations obtained with {\it XMM} \citep{COSMOS_XMM} and {\it
  Chandra} \citep{COSMOS_Chandra}; $3.6\mu$m -- $24\mu$m data from
{\it Spitzer} IRAC and MIPS \citep{COSMOS_Spitzer}, and ground-based
optical and near-infrared photometry from a variety of sources
\citep{COSMOS_oir}.

Our survey is built on three important contributions to COSMOS.  First
are the exquisite photometric redshifts, derived from the photometry of 30 broad, intermediate
and narrow-band filters \citep{COSMOS_photoz}.  These are derived using
a template-fitting technique, and calibrated based on large
spectroscopic samples.  Even for the faintest galaxies in our sample, most of the objects have photometric
redshifts determined to better than 10 per cent (see \S~\ref{sec-specsel}).  Secondly, we use the 10K release of the zCOSMOS
spectroscopic survey \citep{zCOSMOS,zCOSMOS_10K}, from which we obtain
redshifts for moderately bright galaxies ($I_{AB}<22.5$) over most of
the survey area.   This is a sparsely--sampled redshift survey, with
$\sim 40$ per cent sampling completeness.  
\begin{table*}
\begin{tabular}{lllllllll}
Group & RA       & Dec          & $z_{med}$ &  $N_{\rm mask}$&$N_{\rm mem}$ &$\sigma$ &$R_{\rm rms}$&$M_{\rm dyn}$\\
      &\multispan2{\hfil deg (J2000)\hfil} &           && &(km/s) &Mpc&($10^{13}M_\odot$)\\
\hline
40& 150.41991&   1.85265&   0.97225&2&16&765$\pm$124&0.32$\pm$0.03&12.84$\pm$5.50\\
130& 150.02586&   2.20367&   0.93828&3&26&403$\pm$ 48&0.63$\pm$0.06&7.12$\pm$2.40\\
134& 149.65073&   2.20932&   0.94729&3&21&384$\pm$ 60&0.92$\pm$0.07&9.44$\pm$3.69\\
150& 149.97472&   2.31654&   0.93428&4&20&209$\pm$ 31&0.75$\pm$0.07&2.29$\pm$0.90\\
161& 149.95776&   2.34531&   0.94330&4&9&372$\pm$ 56&0.45$\pm$0.07&4.36$\pm$2.01\\
213& 150.39697&   2.49136&   0.87990&2&7&281$\pm$120&0.70$\pm$0.07&3.85$\pm$3.69\\
213a& 150.42715&   2.49992&   0.92650&2&6& 44$\pm$ 17&0.49$\pm$0.10&-\\

\hline
\end{tabular}\caption{Properties of the seven galaxy groups observed with
  GMOS in semester 10A. 
  The position, median redshift $z_{\rm
    med}$, velocity dispersion, and the number of group members
  are determined from our GMOS spectroscopy (combined with available zCOSMOS 10k
  data), as described in the text.  The number of GMOS masks observed
  in each field is given by $N_{\rm mask}$; note that groups 150 and
  161 lie within the same field, as do groups 213 and 213a (a
  serendipitous discovery in the background).  $R_{\rm rms}$ is the rms projected
  distance of all group members from the centre.  The dynamical mass is
  computed as described in the text, from $\sigma$ and $R_{\rm rms}$.
  We do not compute a mass for group 213a, since the velocity
  distribution is unresolved and the membership is poor.\label{tab-gprop}}
\end{table*}

The third pillar of our survey comes from analysis of the deep X-ray
data \citep{COSMOS_alexis}.  Using an established wavelet
technique \citep[e.g.][]{alexis_geec,alexis_sdf} it has
become possible to detect extended X-ray emission from groups of
galaxies out to $z\sim 1$.  Identification of group redshifts is done
using all available spectroscopy, including but not limited to the 10K
zCOSMOS survey.   The combination of X--ray and redshift data provides
a robust catalogue of groups and clusters, to enable follow-up
(Finoguenov et al., in prep).
Masses are estimated from X--ray scaling relations
\citep[e.g.][]{Rykoff2}, and calibrated from a stacked weak-lensing
analysis \citep{COSMOS_lensing}.
Catalogues based on redshift only
\citep[e.g.][]{DEEP2_gerke,zCOSMOS_Knobel} are powerful tools with
which to probe the evolution in the average properties of groups.
However, such catalogues suffer from contamination and projection
effects which make it difficult to interpret studies of individual systems.  
The cross-correlation with X-ray allows one to obtain a more robust
sample of groups.  This potentially introduces a
scientifically interesting bias, if groups with X--ray emission have
fundamentally different populations from groups of similar mass without
such emission.  However, the main difference between optically-- and
X--ray--selected samples seems to be  primarily just that the latter
preferentially selects more massive systems \citep{AEGIS_xray,alexis_geec,2PIGGz_Balogh}.  We thus use this
catalogue as the basis of our group selection, which we describe in the
following subsection.

\subsection{Galaxy group selection}

For our survey, we are interested in the lowest-mass groups that are
robustly identified, and within the redshift range $0.85<z<1$.   We
considered all groups in the Finoguenov et al. (in prep) catalogue,
that lie within this redshift range, have at least 3
spectroscopically-determined members, and were considered secure
identifications.  Specifically, we consider groups that are either
category 1 (good detection, with a well-determined X--ray centre) or
category 2 (secure detection, but with an unreliable X--ray centre).  There are 21 groups satisfying this
selection.    Of these, we have given preference to the lowest--mass,
highest redshift systems, and have avoided targets that already have
$>10$ redshifts used in the identification.  The latter selection is
made in part to avoid massive clusters, and to ensure our spectroscopy increases the total
number of groups in the field with large, spectroscopically-confirmed membership.

Our first observations, described in \S~\ref{sec-gemini}, targeted six
of these groups.  Their properties are tabulated in
Table~\ref{tab-gprop}.

\subsection{Gemini Observations}\label{sec-gemini}
\subsubsection{Spectroscopic target selection}\label{sec-specsel}
A crucial part of our strategy is the use of photometric redshifts to
select potential group members.  We give highest priority to galaxies
that have $21.5<r<24.75$, and a redshift within 2$\sigma_{zphot}$ of the estimated group
redshift, where $\sigma_{zphot}$ is the 68 per cent confidence level on
the photometric redshift\footnote{At our redshift of interest, the average
uncertainty on $z_{\rm phot}$ increases from $\sigma_{zphot}\sim 0.007$ for
the brightest galaxies to $\sigma_{zphoto}\sim 0.04$ for those at our limit of
$r=24.75$.  Even for these faintest objects, 90 per cent of the
galaxies have $z_{\rm phot}$ uncertainties of less than $\sigma_{zphot}<0.07$.}.
Secondary priority slits are allocated to galaxies
with $15<r<24.75$ and $0.7<z_{\rm phot}<1.5$. 

We designed 3--4 GMOS masks on each target, using the {\sc gmmps}
software.
On each mask we are able to allocate $40$--$50$ slits.  Typically, well
over half the slits on the first mask are allocated to our top priority
objects.  This fraction decreases on subsequent masks, as we use up
these targets.  In most cases, with three masks we are able to target
at least 40 of these high priority galaxies. All masks for a given
target use the same alignment stars and 
are at the same position angle; thus a small fraction of the CCD area is unusable
regardless of the number of masks obtained.

\subsubsection{Observations}
We obtained 40 hours worth of observations on Gemini South, in the Band
1 queue during semester 10A.  This returned science data of 2 hours
on-source exposure, for 14 masks in 5 fields, covering six groups (two
of the groups are located within the same GMOS field).  All
science observations were obtained in clear conditions with seeing 0.8
arcsec or better in $i$.    The number of masks actually observed is
given in Table~\ref{tab-gprop}.

The spectroscopy was obtained in nod \& shuffle mode \citep{GBH},
nodding the telescope by $\pm$0.725 arcsec from the centre of the slit,
every 60s.  This places the galaxy at the lower end of the 3 arcsec slit half
the time, and at the upper end the other half of the time.  We used
microshuffling mode, so charge was shuffled by 21 pixels (3.05 arcsec).  Four
exposures of 30 min each were taken, with small dithers in the spectral
direction for the purpose of removing bad pixels and interpolating over
the chip gaps.

Slits were 1 arcsec wide, and we use the R600 grism with OG515
order blocking filter.  The detector was binned $2\times2$, resulting
in a pixel scale of 0.146 arcsec/pix, and a dispersion of
$0.93$\AA/pix.  The resulting spectral resolution is limited by the
slit width, $\sim 6.4$\AA.  

Science observations were interspersed with GCAL flats.  Calibration
exposures for each mask consisted of CuAr arc lamp observations and
twilight flats.  In addition, we observed the standard star LTT6248
with a 1 arcsec, long slit, for flux calibration.

\subsubsection{Data reduction}\label{sec-dr}
All data were reduced in {\sc iraf}, using the {\sc gemini} packages
with minor modifications.  Slits were first identified from the GCAL flat
exposures.   A bad pixel mask was created using the routine {\sc gbpm},
and a set of six long- and three short-exposure image flats.  Pixels in
the chip gaps, and outside the spectroscopic field of view, are also
marked as bad pixels.

A bias frame was subtracted from all science data.  Sky
subtraction is done simply, using the {\sc gnsskysub} routine, by
subtracting from each science frame a copy of itself, shifted by 21
binned pixels.  This results in a positive-flux object spectrum in one
half of the slit, and a negative-flux spectrum in the other half.

To generate a noise vector we use the same technique, but {\it add} the
shifted images.  This gives pixels, at the location of the galaxy, that
include object flux and twice the sky flux, which is equal to the
expected Poisson variance.  We add to this twice the readnoise squared,
to obtain the final variance spectrum.

Each sky-subtracted science image is then cleaned of cosmic rays,
flat-fielded, and cut into individual slits.  Corresponding arcs are
extracted and wavelength calibrated; this calibration is then applied
to each science frame.  The four exposures are added, after applying a
shift to account for the spectral dither and ignoring bad pixels
identified in the bad pixel mask, which has been propagated through the
same reduction procedures as the science frame.  The same is done for
the ``noise'' frame described above.

The negative-flux spectrum is then subtracted from its positive
counterpart, for each spectrum, using by default a 3.5 pixel wide
aperture (this is tweaked in a few cases, where necessary).  A wavelength-dependent sensitivity function is determined by
extracting the standard star spectrum, and comparing the flux to
tabulated values.  This is applied to the final 1D and 2D object
spectra, to give flux-calibrated spectra.

Our final step is to correct for telluric absorption.  This is of
particular importance since the H\&K absorption lines in galaxies at
our target redshift $z\sim 0.95$ lie right on top of the A-band
telluric line.  We use our standard star observation to extract
spectral regions around A-band ($7500<\lambda/$\AA$<7700$) and B-band
($6850<\lambda/$\AA$<6900$).  A smooth continuum is subtracted, and we
use the {\sc IRAF} task {\sc telluric} to apply the correction to each
spectrum according to the airmass in the image header.

\subsubsection{Redshift Determination}
Redshifts were measured by adapting the {\sc zspec} software, kindly
provided by R. Yan, used by the DEEP2 redshift survey \citep{DEEP2,Davis+07}.  This
performs a cross-correlation on the 1D extracted spectra, using linear combinations of template
spectra.  The corresponding variance vectors described in
\S~\ref{sec-dr} are used to weight the cross-correlation.  

We use templates of absorption line and emission line galaxies only,
and for each galaxy {\sc zspec} returns a short list of possible
redshifts, with associated $\chi^2$ and $R$ \citep{TD79} values.  Every
spectrum is visually inspected, both in the 1D and 2D format.
Particularly valuable is inspecting the 2D image prior to combining the
positive- and negative-flux objects.  Real emission lines are clearly
identified by their dipole signature, and false peaks due to cosmic
rays or other effects are easily removed.  Generally, the 1D spectra
are boxcar smoothed to $\sim 10$\AA\ pixels for visual inspection.

We adopt a simple, four-class method to quantify our redshift quality.
Quality class 4 is assigned to galaxies with certain redshifts.
Generally this is reserved for galaxies with multiple, robust
features.  With our (unbinned) spectral resolution, we are able to just
resolve the [OII] doublet; in this case, a clear detection of the doublet alone
would warrant a class 4 redshift.  Quality class 3 are also very
reliable redshifts, and we expect most of them to be correct.  These
include galaxies with a good match to Ca H\&K for example, but no
obvious corroborating feature.  Similarly, single emission lines where
a doublet is not convincing are generally assumed to be [OII] and given
quality class 3.  We would also use this in cases where H\&K are
detected in a region of telluric absorption, but there is at least one
other likely match to an absorption feature.  We take particular care
not to assign class 3 or 4 to a galaxy for which H\&K are the only
identifiable features, and lie on a telluric absorption line.

Class 2 objects correspond to ``possible'' redshifts.  These include
some spectra that are reasonably likely to be correct (e.g. H\&K on top
of a telluric line and no other corroborating features); but also some
that are little more than guesses.  Class 1 is reserved for ``junk'',
with no chance of obtaining a redshift.

In this analysis we only consider galaxies with class 3 or 4 quality redshifts.
\begin{figure}
\leavevmode \epsfysize=8cm \epsfbox{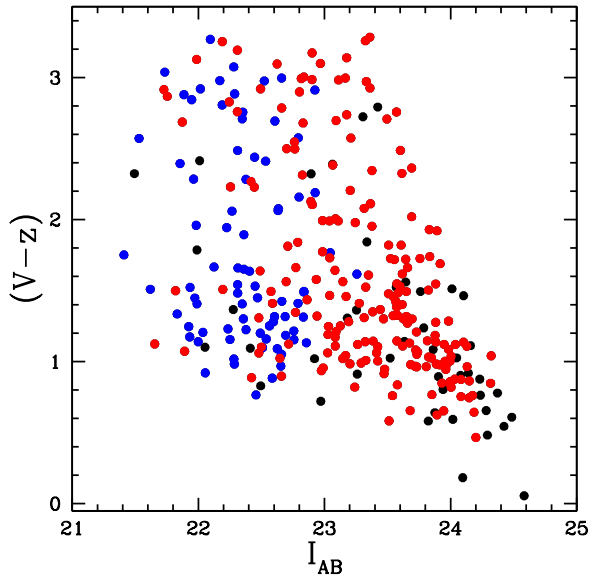} 
\caption{The colour-magnitude distribution for all galaxies with
  spectroscopy (from either GEEC or zCOSMOS) that lie within our GMOS fields, and
  have $0.8<z_{\rm phot}<1.5$, within their 1$\sigma$
  uncertainties.  {\it Red points}
represent GEEC targets with secure redshifts (quality 3 or 4), while blue points
indicate secure zCOSMOS 10k redshifts (quality $>2$).   The remaining,
{\it black points}, are therefore those with a spectrum and  $0.8<z_{\rm phot}<1.5$, but without a secure 
redshift.   \label{fig-cmag_z} }
\end{figure}

\subsection{zCOSMOS Observations and final catalogue}\label{sec-cat}
We also include in our analysis redshifts from the 10K release of
zCOSMOS \citep{zCOSMOS_10K}.  We include all galaxies with redshift quality greater
than 2.0, which have a high probability ($>90$ per cent) of being
correct (note the zCOSMOS quality flags are defined differently from
our own).  These provide a nice complement to our observations, by
design, as they are restricted to the brighter galaxies that are not
given priority in our target selection.

Figure~\ref{fig-cmag_z} shows our final sample in colour-magnitude
space.  All optical colours and magnitudes are computed within a
3\arcsec\ diameter aperture, on psf--matched images, as described in \citet{COSMOS_oir}.
The galaxies plotted here are all those with an available spectrum,
lying within our GMOS
fields, and with $0.8<z_{\rm phot}<1$ within
their 1$\sigma$ uncertainties.  We choose the $(V-z)$ colour as it
brackets the 4000\AA\ break at $z\sim 0.95$,
and the $I_{AB}$ magnitude for comparison with zCOSMOS.  Red points
represent GEEC targets with secure redshifts, while blue points
indicate zCOSMOS 10k galaxies with secure redshifts.  Distinct sequences of red and
blue galaxies are apparent, and our
redshift success rate is very high at all colours (see
\S\ref{sec-zsuc}).  Our GEEC spectra probe {\it red } galaxies up to 1.0 mag fainter
than found in the 10k survey ($I_{AB}<22.5$), 0.65 mag fainter than
DEEP2 \citep[$r<24.1$]{DEEPII_envt_again} and $\sim 0.25$ mag fainter
than EDisCS at this redshift \citep[$I_{\rm Vega}<23$, or $I_{\rm AB}\lesssim
23.3$]{Halliday+04}.  For blue galaxies we gain additional $\sim 0.5$
mag depth
relative to the $I-$selected zCOSMOS and EDisCS surveys.

\section{Analysis}\label{sec-anal}
\subsection{Spectroscopic and Redshift completeness}\label{sec-zsuc}
We first consider the sampling completeness of our GMOS targets,
defined as the fraction of potential targets for which we actually
obtained a spectrum.   Here, we
define the complete sample to be all priority 1 galaxies (i.e. all
$21.5<r<24.75$ galaxies that have a photometric redshift within 2$\sigma$ of the
estimated group redshift) that are within the GMOS field of view.  Note
that 100 per cent completeness in this sense could never have been achieved,
because some of the field of view is permanently inaccessible due to
the acquisition stars and, in a few cases, guider arm.  
We characterize our sampling completeness in terms of the IRAC [3.6$\mu$m]
magnitude, since this is more closely related to galaxy stellar mass
than the $r$ band magnitude on which our selection was based.  We
include galaxies from the zCOSMOS 10K spectroscopic catalogue here as
well, which mostly includes galaxies brighter than $I_{AB}=22.5$.
Although our target
selection does not explicitly depend on galaxy colour, the reliance
on photometric redshift uncertainty potentially introduces a
colour-dependence.  Thus we consider our sampling completeness as a
function of both [3.6$\mu$m] magnitude and $(V-z)$ colour.
Specifically, we divide this colour-magnitude plane into several bins
and calculate the completeness in each bin.  For each colour bin we
define a completeness function by fitting 
a line to the completeness as a function of [3.6$\mu$m] magnitude.  We find that, regardless
of how we do this, our
completeness is remarkably uniform; it ranges from $\sim 0.6$ to $\sim 0.75$ with little
dependence on colour or magnitude.  This may be partly because
the sample is too small yet to define the selection function with high
enough precision, and we will revisit this when the survey is
complete.  

\begin{figure}
\leavevmode \epsfysize=8cm \epsfbox{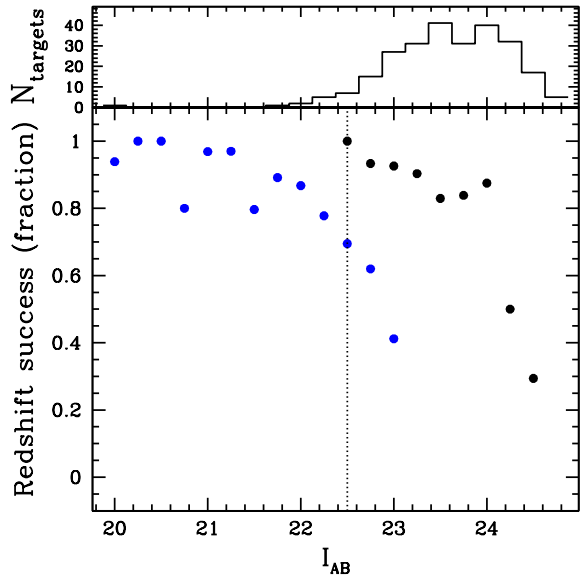} 
\caption{{\it Top:} The histogram shows the number of priority 1, GEEC
  spectra, as a function of $I_{AB}$ magnitude. {\it Bottom:} The
  redshift success rate of GEEC, defined as the fraction of priority 1
  galaxies with good (quality 3 or 4) redshifts, is shown as the {\it
    black points}.  The {\it blue} points represent a similar quantity
  for the zCOSMOS 10k sample within our field; here it is the fraction
  of all galaxies in our GMOS fields with a
  zCOSMOS spectrum that have a redshift quality of $>2.0$.  The
  vertical, dotted line at $I_{AB}$ represents the nominal, zCOSMOS
  completeness limit.  With GEEC, we are highly complete for galaxies
  up to 1.5 mag fainter than this.\label{fig-zsuccess0} }
\end{figure}

Another potential source of incompleteness is failure to obtain
redshifts for targeted galaxies.  We characterise our redshift success
rate as a function of $I_{AB}$ magnitude, since this corresponds most
closely to the wavelength of features (like Ca H\&K) commonly used for
redshift identification at $z\sim 0.9$.  This success rate, shown in
Figure~\ref{fig-zsuccess0}, is defined as
the fraction of priority 1, targeted galaxies for which we obtained a redshift with
quality 3 or 4.  We find a remarkably high success rate of $>80$ per
cent, for $I_{AB}<24$.  This is a testament in part to the high-quality
spectra obtained at Gemini, and in particular the success of the nod \&
shuffle technique, which results in near-perfect sky subtraction in a
spectral region dominated by sky emission lines.  Moreover, we are
helped by the exquisite photometric redshifts, which means our priority
1 list contains little contamination from higher redshift galaxies for
which redshifts are difficult to obtain.

We show, as the blue points, the redshift success of the zCOSMOS 10k
sample, restricted to the regions covered by our GMOS observations.
Here the success rate is defined as the fraction of secure (quality
$>2$) redshifts from all zCOSMOS targets within our GMOS fields of view. The success rate is also very high here, $>70$ per cent for
$I_{AB}<22.5$, which is their nominal completeness limit.  
We have also checked for a colour
dependence on these success rates, and find no significant trend, as is
visually apparent on Figure~\ref{fig-cmag_z}.
Thus, both zCOSMOS and GEEC are highly successful at obtaining secure
redshifts even for red, absorption-only galaxies.  Many of these
``failures'' are in fact likely to be higher redshift galaxies,
in which case they do not affect the success rate in our limited
redshift range of interest.  
We therefore apply no further corrections for redshift incompleteness;
the only weight comes from the sampling fraction.

\begin{figure}
\leavevmode \epsfysize=8cm \epsfbox{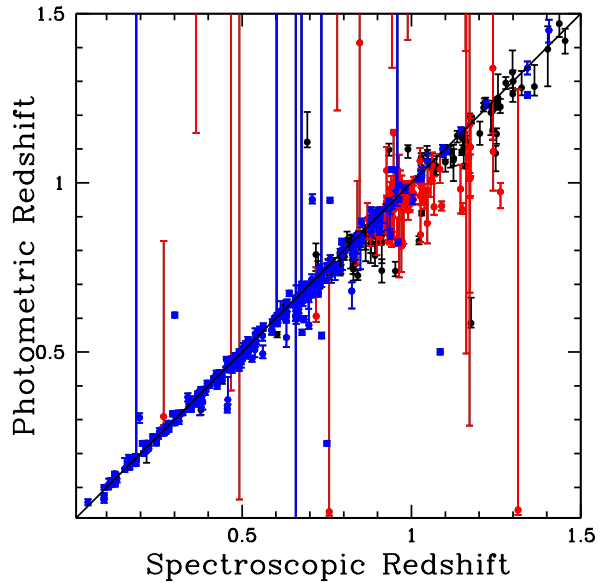} 
\caption{Spectroscopic redshifts are compared with photometric
  redshifts and their uncertainties.  All
  galaxies with a secure redshift are shown.  {\it Red points} are those
  GEEC galaxies with priority 1, which are
  those that have $z_{\rm phot}$ within 2$\sigma$ of the group
  redshift; {\it black points} represent GEEC galaxies with lower priority.  {\it Blue points} are zCOSMOS 10k spectra within the same
  fields.  Our spectra greatly increase the completeness and depth of
  the 10k survey at the redshift of these groups.  \label{fig-cfz} }
\end{figure}
\begin{figure}
\leavevmode \epsfysize=8cm \epsfbox{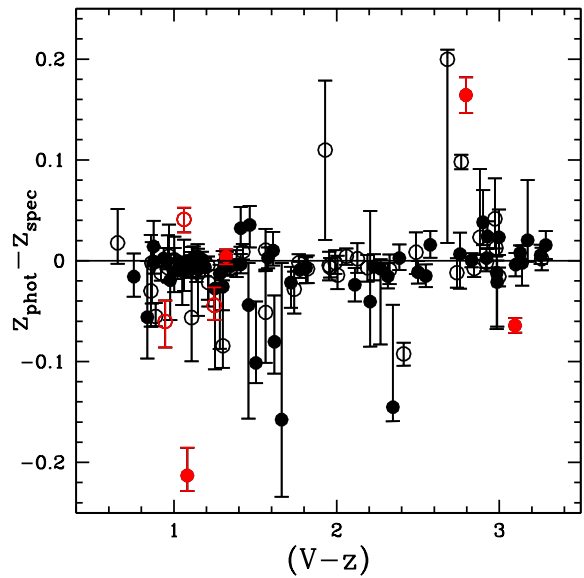} 
\caption{For confirmed group members with $0.7<z_{\rm phot}<1.5$ we show the
  difference between spectroscopic and photometric redshifts, as a
  function of $(V-z)$ colour.  
  Asymmetric errorbars are 68 per cent confidence intervals on $z_{\rm
    phot}$.  {\it Filled symbols} are quality class 4, which means the
  spectroscopic redshift is certain, while {\it open symbols} are class
  3 and thus generally reliable (and used throughout this analysis).
  The {\it red points} identify those few group galaxies that were
  allocated lower priority during mask design because their $z_{\rm
    phot}$ is more than 2$\sigma$ away from the mean group redshift.
  These make up only $\sim 6$ per cent of the total group population,
  with no obvious colour dependence.\label{fig-cfz3} }
\end{figure}
In Figure~\ref{fig-cfz} we compare our spectroscopic redshifts with the
corresponding photometric redshift and its uncertainty.  Red points
represent secure redshifts for priority 1 galaxies in GEEC.  These all,
by definition, have a $z_{\rm phot}$ within 2$\sigma_{zphot}$ of the group
redshift ($0.88<z<0.97$ for this sample).   Effectively, this selects
galaxies that either have $z_{\rm phot}$ close to the group redshift,
or galaxies with poorly determined $z_{\rm phot}$, which have large
error bars.
We see that not only do the
spectroscopic and photometric redshifts agree very well, but that our
preselection allows us to be efficient at targeting potential
group members.  The blue points represent secure redshifts in the
zCOSMOS 10k sample.  

During mask design, we preferentially target galaxies with photometric
redshifts within $2\sigma$ of the mean group redshift.  Since we
account for the uncertainty in $z_{\rm phot}$, we do not expect to be
strongly biased against galaxies for which $z_{\rm phot}$ is poorly
determined due, for example, to a lack of strong features in the
spectrum.  However, we are potentially biased against  ``catastrophic
failures''.   To address this, in Figure~\ref{fig-cfz3} we show the
difference between the spectroscopic and photometric redshift for all
confirmed group members (see Section~\ref{sec-gmem}) with secure
spectroscopic redshifts and $0.7<z_{\rm
  phot}<1.5$, as a function of their $(V-z)$ colour.  Red points
identify ``priority 2'' galaxies, which are those lower priority
targets with  $z_{\rm phot}$ more than 2$\sigma$ away from the group
redshift.  These make up only $\sim 6$ per cent of the sample, with no
measureable colour dependence.   There is also no strong dependence on the
quality of the spectroscopic redshift; the fraction of priority 2
galaxies that are actually group members is similar amongst quality 3
and quality 4 redshifts.  We conclude therefore that our $z_{\rm phot}$
preselection does not introduce a large bias, though we will be able to
test this more robustly at the end of the survey.

At this early stage of our survey, we have few duplicate observations
with which to check our spectroscopic redshifts and determine a robust
uncertainty.  We do have eight good quality redshifts for galaxies that
have an existing, good quality zCOSMOS redshift.  From these, there
appears to be a small bias in the sense that our redshifts are smaller
by $7.72\times10^{-4}$, corresponding to a rest-frame velocity shift of
$\sim 120$ km/s at the redshift of interest.  Although this should be
treated as a preliminary offset, we correct our redshifts for it here.
We use the same eight galaxies to estimate our redshift uncertainty and
find $\sigma_z \sim 4.1\times10^{-4}$, corresponding to a rest-frame velocity
uncertainty of $\sim 65$km/s at $z=0.9$.

\subsection{Stellar mass measurements and k-corrections}
We fit the spectral energy distribution for each galaxy, using all
available photometry, following the method described in \citet{GEEC_sedfit}.  Briefly,
this is done by comparing with a very large grid of template
\citet{BC03} models, covering a wide range of parameters, assuming a
\citet{Chab} mass function.  The
procedure is based on that of \citet{Salim07}, and assumes the star
formation history of a galaxy can be represented by an exponential
model with superposed bursts.
\begin{figure}
\leavevmode \epsfysize=8cm \epsfbox{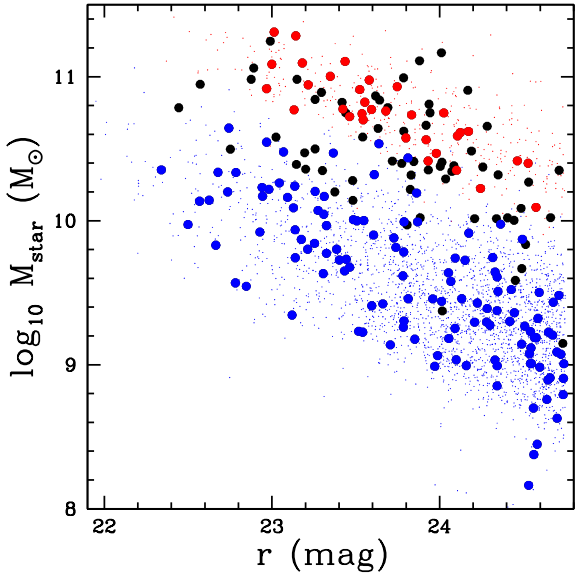} 
\caption{The stellar masses, computed for a \citet{Chab} mass function
  and based on \citet{BC03} models, are shown as a function of $r$
  magnitude for a random subset of the photometric field sample (small points) and our
  spectroscopic sample (large, filled points), for $0.8<z<1.0$.  The
  reddest galaxies,  $(V-z)^{0.9}>2.7$, are shown as red points, while
  the bluest galaxies  $(V-z)^{0.9}<1.5$ are shown in blue (the
  remaining black
  points are those of intermediate colour).  Our
  selection limit of $r<24.75$ imposes a colour--dependent mass limit,
  of $M_{\rm star}<10^{10.1}M_\odot$ for red galaxies, and
  $M_{\rm star}<10^{8.8}M_\odot$ for the bluest. \label{fig-smass_r} }
\end{figure}

Stellar mass estimates are quite robust to the details of the fits, in
part because of the availability of {\it Spitzer} IRAC data.  The main
assumptions that could lead to systematic errors are the initial mass
function (IMF), the dust model \citep[we use ][]{CF00},  and the choice of galaxy evolution model.  Of particular
concern could be the omission of thermally-pulsing AGB stars, which become increasingly
important at higher redshifts \citep{Maraston}.  This will be more
relevant when considering comparison with $z=0$ observations, which we
defer to after completion of the survey.  For our purposes here, we are
interested in comparing group galaxies with their field counterparts.
Systematic effects like this on the mass estimates are not likely to be
very different for the two populations.  Thus we expect our conclusions
to be robust to this and similar assumptions about IMF and dust geometry.

We k-correct all colours to $z=0.9$, the redshift of interest for our
survey, using the {\sc kcorrect} IDL software of \citet{kcorrect}.
We denote these colours as, for example,  $(V-z)^{0.9}$.
 
In Figure~\ref{fig-smass_r} we show our derived stellar masses as a
function of $r$ magnitude (our selection band), for both the
photometric field sample (small points) and our spectroscopic sample,
within $0.8<z<1.0$.  The points are separated into the bluest
galaxies, $(V-z)^{0.9}<1.5$, and the reddest galaxies,
$(V-z)^{0.9}>2.7$.  Our selection limit of $r<24.75$ implies we are
100 per cent complete above a mass limit of  $M_{\rm
star}\gtrsim10^{10.5}M_\odot$; however, we are mostly complete for $M_{\rm
star}\gtrsim10^{10.1}M_\odot$, missing only some of the reddest
galaxies.  We will therefore treat this as our nominal completeness
limit throughout the paper.  Note that the
blue galaxies are complete for $M_{\rm star}\gtrsim10^{9.6}M_\odot$,
and we probe masses as low as  $M_{\rm star}\lesssim10^{8.8}M_\odot$.

\subsection{Group masses and membership}\label{sec-gmem}
For each group, we inspect the spatial and redshift distribution of the
galaxies.  Full results will be presented for all groups, after
completion of the survey.  

We start by considering all galaxies within $r=1$ Mpc of the
nominal X--ray group centre from Finoguenov et al. (in prep).  The velocity dispersion
of these galaxies is determined using the gapper estimate
\citep{Beers}.  We also compute the (unweighted) mean
spatial position of the galaxies, and the {\it rms} projected
separation from this centre, which we call $R_{\rm rms}$.   We then
iterate this process, typically clipping galaxies with
velocities $>1.5\sigma$ from the median redshift, and positions
$r>2R_{\rm rms}$ from the recomputed centre\footnote{The clipping
  parameters were tweaked for group 161, which is very  close to group
  150 in redshift and position.  For this group we exclude galaxies
  with velocities $>1.4\sigma$ from the median redshift, and positions
$r>1.6R_{\rm rms}$ from the recomputed centre.  The properties
(dynamical and stellar mass) of this particular group are quite sensitive to this choice.}.    This
converges after a few iterations, and we adopt the final
$\sigma$ and $R_{\rm rms}$.  Group members are then defined as those within
$3\sigma$ of the median redshift, and within $2R_{\rm rms}$ of the centre.
Uncertainties on $\sigma$ and $R_{\rm rms}$ are
then computed using a jackknife method, iterating only over this list
of group members.  Thus, these uncertainties do not include
systematic uncertainties due to the clipping process, which are likely
to be at least $\sim 15$ per cent \citep[e.g.][]{CNOC2_groupsI}.

\begin{figure}
\leavevmode \epsfysize=8cm \epsfbox{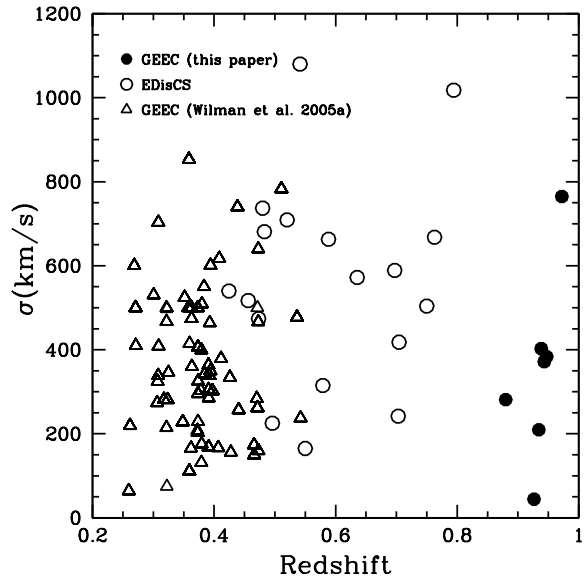} 
\caption{The velocity dispersion, as a function of redshift, for three
  samples of groups and clusters with high spectroscopic completeness.
  The present group sample is shown as {\it filled circles}, while our
  lower-redshift GEEC sample is shown with {\it open triangles}.  The
  {\it open circles} represent the groups and clusters of the EDisCS
  survey \citep{Halliday+04}.\label{fig-ediscs} }
\end{figure}

We show how our sample compares with other, related surveys, in
Figure~\ref{fig-ediscs}.  Points show the velocity dispersion of each
group, as a function of redshift, for our present survey (filled
circles), our previous GEEC sample at $0.3<z<0.6$ (open triangles), and
the EDisCS cluster and group sample \citet[][open
circles]{Halliday+04,Pogg05}.  The median redshifts,
velocity dispersions, and number of members in each of our groups is
given in Table~\ref{tab-gprop}.

We estimate a dynamical mass, from the velocity dispersion $\sigma$ and
the radius $R_{\rm rms}$,
\begin{equation}
M_{\rm dyn}=\frac{3}{G}R_{\rm rms}\sigma^2.
\end{equation}
The uncertainty is determined by propagating the jackknife
uncertainties on $\sigma$ and $R_{\rm rms}$.  The factor $3$ in this
equation is based on the assumption of isotropic orbits and an
isothermal potential, but is only weakly dependent on those assumptions \citep{LM01}.

%

Groups 150 and 161 are located within the same GMOS field.  It is clear
that group 161 is dynamically and spatially separated from group 150,
but they are very close, separated by less than $1$ Mpc and $1000$
km/s (rest frame).  Thus they may be part of the same, interacting
system.  The assignment of members to one group or the other is not
entirely obvious, and our clipping algorithm was tuned to achieve a
reasonable--looking separation.  

Finally, the membership of group 213 is quite poor, and our completeness in this field is worse than elsewhere.  We did, however,
detect a serendipitous group (named 213a) at higher redshift, $z=0.9254$.  This
group has six members, and a nominal velocity dispersion of only
$44\pm17$ km/s, rest frame --- smaller than would be expected from
redshift uncertainties alone, and likely dominated by systematic
uncertainty.  The group's position is significantly offset from the
X--ray detection, which is likely still correctly identified with the
lower redshift system.  We include both groups in our analysis;
however, we do not attempt to determine a dynamical mass of group 213a
since the velocity dispersion is clearly unresolved.
 
Our observations confirm that these are all low-mass systems, as expected
from their X--ray fluxes.  The dynamical masses of the targeted
groups range from
$2.3\pm0.9\times10^{13}M_\odot$ to $1.3\pm0.5\times10^{14}M_\odot$.

\section{Results}\label{sec-results}
\subsection{Dynamics and stellar fractions}
\begin{figure}
\leavevmode \epsfysize=8cm \epsfbox{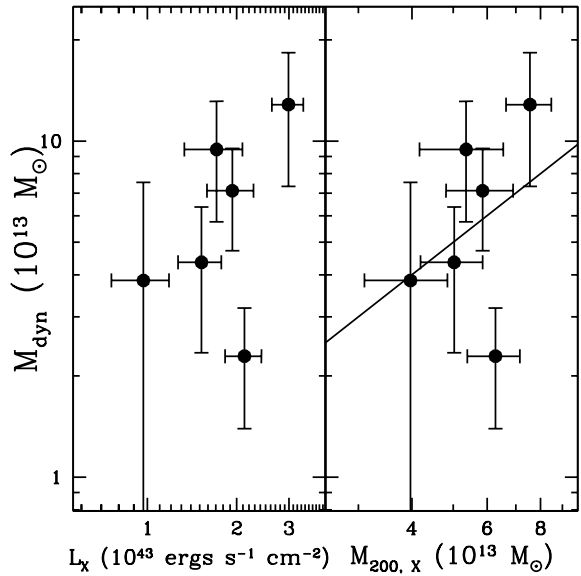}
\caption{{\it Left:} The correlation between $M_{\rm dyn}$ and X--ray
  luminosity, $L_X$, is shown for the six targeted groups in our
  sample (i.e. excluding 213a, which is not detected in X--rays).  {\it
    Right: } Here we show the same data, but as a function of $M_{\rm
    200, X}$ as determined directly from $L_X$ assuming a scaling
  relation and calibrated from  stacked weak lensing analysis
  \citep{COSMOS_lensing}.  
\label{fig-mdyn_lx_mx200} }
\end{figure}

With at least $\sim 15$ spectroscopic members for most groups, we are
able to make reasonable estimates of the dynamical mass, as described
in \S~\ref{sec-gmem}.  In Figure~\ref{fig-mdyn_lx_mx200} we compare
these masses with the X--ray luminosity (left panel) and with the
virial mass $M_{200,X}$ estimated from this luminosity (right panel).
$M_{200,X}$ is estimated assuming a scaling relation and calibrated based on a
 stacked weak lensing analysis \citep{COSMOS_lensing}.  

The
 dynamical mass uncertainties are generally large, but the masses agree remarkably well
 with those estimated from the X--ray emission.  The one outlier, which
 has a relatively low dynamical mass for its X--ray emission, is group
 150; this group is close to and likely
associated with 161.  Its position on this plot is quite sensitive to
the precise membership assigned; that is, the systematic uncertainty
due to the choice of clipping parameters are likely dominant over the
statistical uncertainties. 

Total stellar masses can often be
determined with much greater precision, with the dominant statistical noise term
coming from the uncertainty in the radius $R_{\rm rms}$ within which the stellar mass is
summed.   The total stellar mass is computed over all galaxies in the
sample; since we are incomplete for red galaxies with
$M<10^{10.1}M_\odot$, this is actually a lower limit.  For the blue
galaxies, about 75 per cent of the stellar mass is at
$M>10^{10.1}M_\odot$.   Assuming this to be true for the red population
as well, we estimate that our stellar masses underestimate the true
total by less than 10 per cent.  The statistical uncertainty is
calculated from the jackknife uncertainties in $R_{\rm rms}$ and
$\sigma$.  In several cases, the group members with $r<R_{\rm rms}$ and
$|\Delta z|<3\sigma$ do not change within the uncertainties on these quantities, which
leads to a stellar mass estimate with, formally, zero uncertainty.  We
therefore impose an arbitrary, minimum uncertainty of 10 per cent on
total stellar masses.

\begin{figure}
\leavevmode \epsfysize=8cm \epsfbox{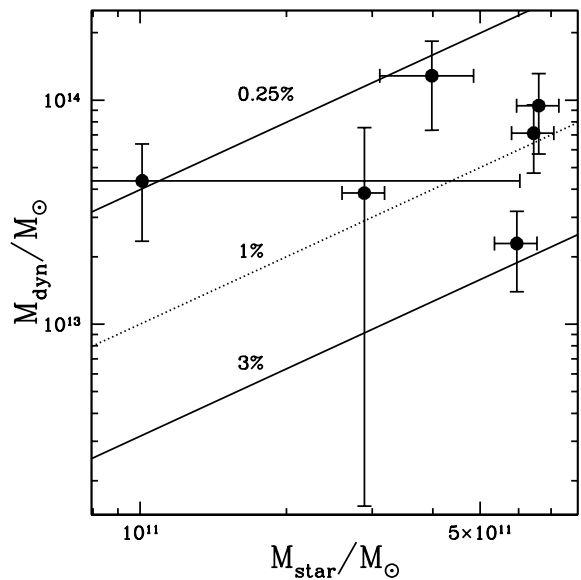}
\caption{We show the measured dynamical mass as a function of total stellar mass, for each of our
  groups except the
  serendipitous group 213a.  The stellar mass is the sum of all
  spectroscopically confirmed group members within a radius $R_{\rm
    rms}$.  The dotted line indicates a constant stellar fraction of 1
  per cent; the solid lines on either side represent fractions of 0.25
  per cent and 3 per cent, as indicated.   
  \label{fig-mdyn_mstar} }
\end{figure}

In Figure~\ref{fig-mdyn_mstar} we show the relationship between
dynamical and stellar mass, for all six targeted groups.   We exclude the
serendipitous group 213a, since the velocity dispersion is unresolved;
with only six members we cannot determine a reliable dynamical mass.
For the remainder, the stellar fractions
range from $\sim 0.25$ per cent to $\sim 3$ per
cent, similar to what is observed both locally
\citep[e.g.][]{2PIGGz_Balogh} and at higher redshift \citep{G+09}.
The group with the highest stellar fraction is 
group 150 which, as mentioned previously, has a larger systematic
uncertainty on its dynamical mass due to its close proximity to 161.
The membership of group 161 itself is very sensitive to the choice of $R_{\rm rms}$, and the
uncertainty in this quantity leads to a large uncertainty in stellar mass.
Thus, although there are hints of
real diversity in the stellar fraction amongst our groups, the sample
is too small at this time to draw any strong conclusions.
\begin{figure*}
\leavevmode \epsfysize=8cm \epsfbox{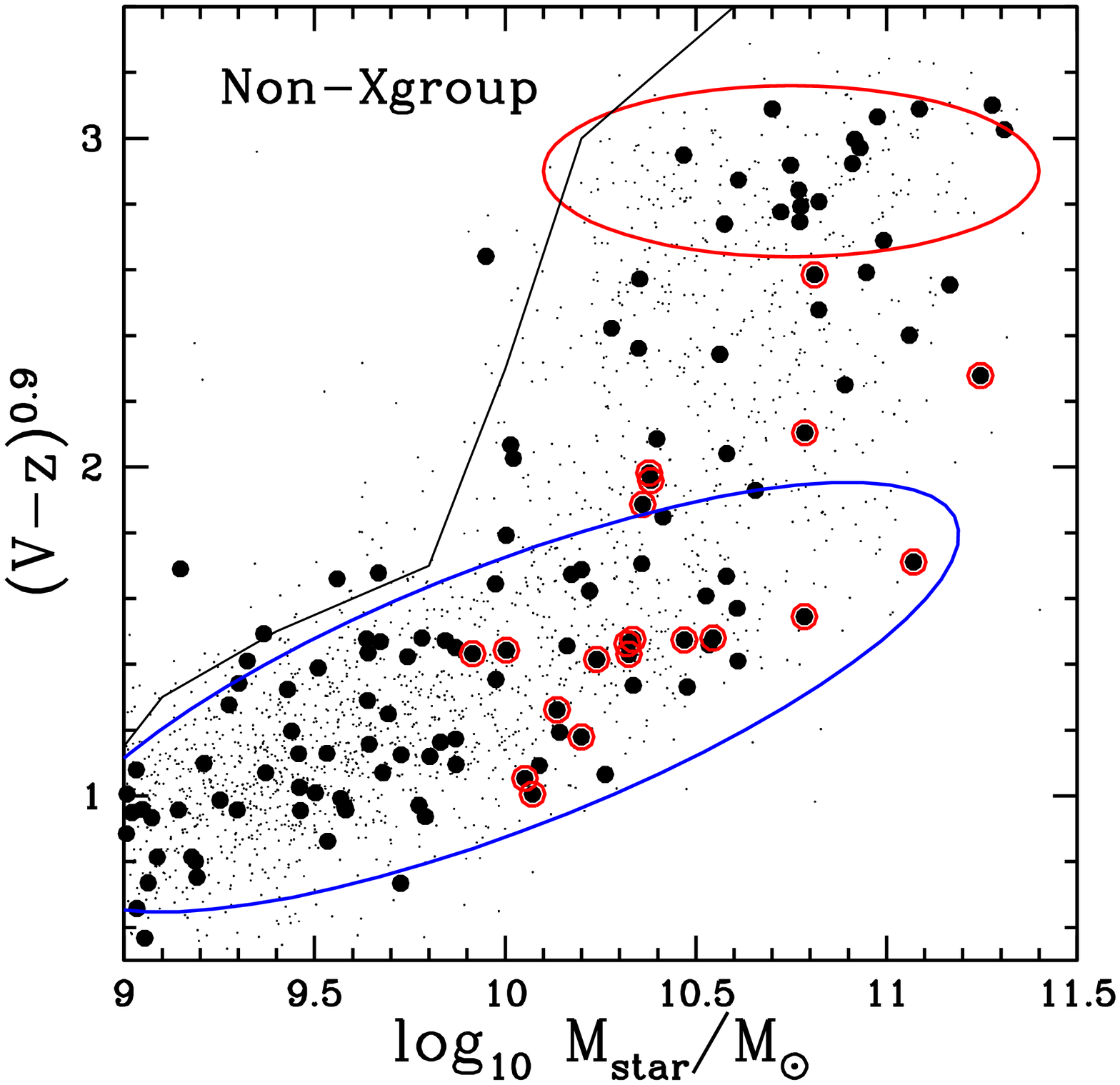} \leavevmode \epsfysize=8cm \epsfbox{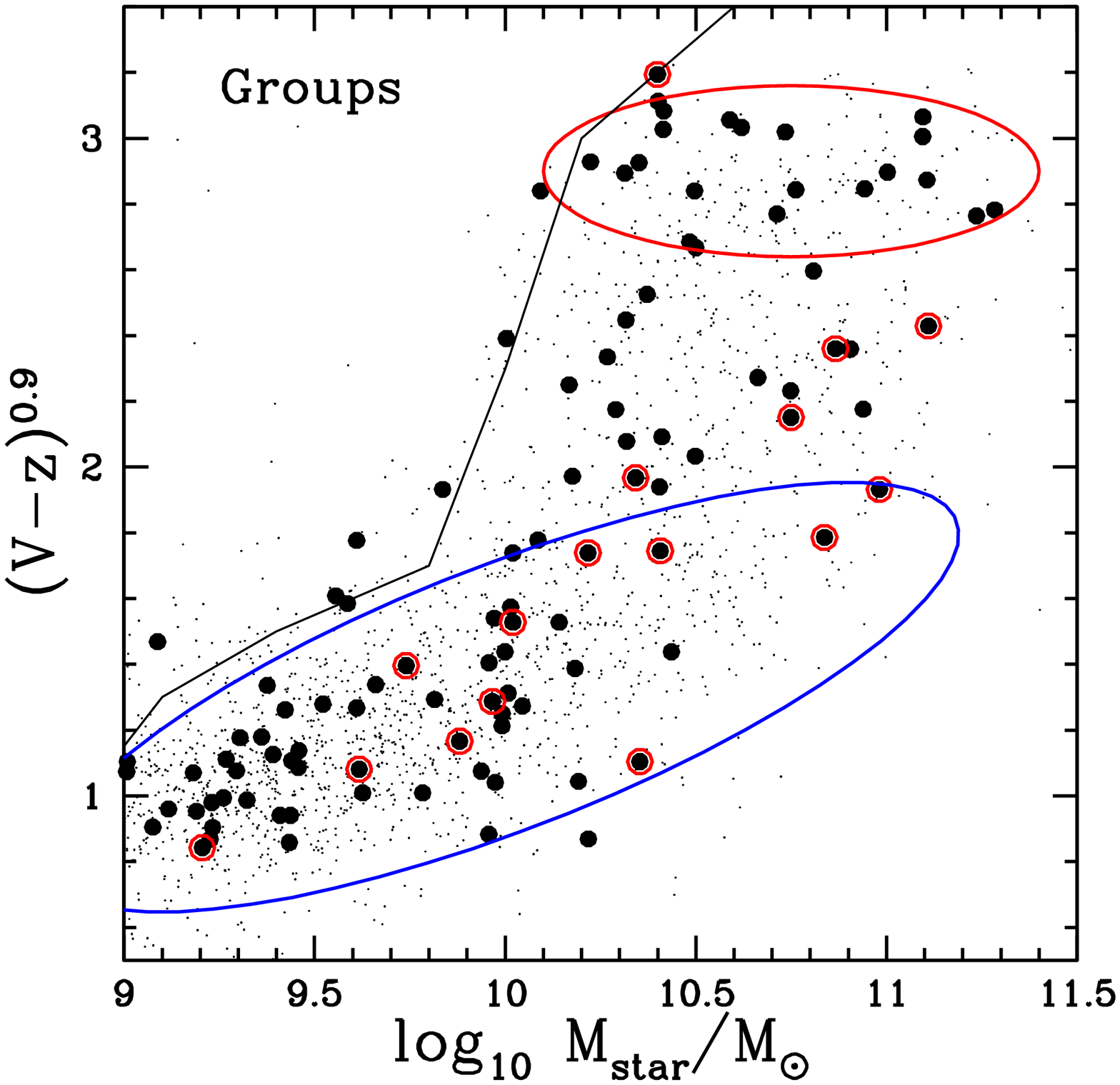}
\caption{{\it Left: }The colours (k-corrected to $z=0.9$) and stellar masses of
  galaxies are shown for two samples.  The small, black points are
  galaxies from the full COSMOS photometric catalogue, with $r<24.75$
  and $0.8<z_{\rm phot}<1.0$.  We only show a random 10\% of the points, for
  legibility.  This represents the parent
  distribution from which our spectroscopic surveys (over a much
  smaller field) are drawn.  The large, filled points represent
  galaxies with secure, spectroscopic redshifts at $0.8<z<1$, that lie within our GMOS fields
  but are not assigned
  to a group in our survey.  We refer to these as ``non-Xgroup'' environment
  galaxies, to distinguish them from the general field (represented by
  the small points) which contains galaxies in all environments.  The points circled in red
indicate galaxies with 24$\mu$m detections. Note the absence of low-mass, red-sequence galaxies in the
  non-Xgroup galaxies, and the dominance of the blue sequence for
  $M_{\rm star}<10^{10.6}M_\odot$.  For illustration purposes only, the
  large red and blue ellipses approximately identify the ``red
  sequence'' and ``blue cloud'' populations.  The solid line indicates
  our approximate completeness limit imposed by the $r>24.75$
  selection.
{\it Right: }The same, but now the filled
  points correspond to group members.  Here, the faint end of the red
  sequence, visible in the parent photometric redshift catalogue, is
  filled in with confirmed group members.  Moreover, there appears to
  be a prominent third population, that lies between the blue and red sequences
  at $(V-z)^{0.9}\sim 2$.\label{fig-cmass_z} }
\end{figure*}

\subsection{Galaxy populations}
\begin{figure*}
\leavevmode \epsfysize=8cm \epsfbox{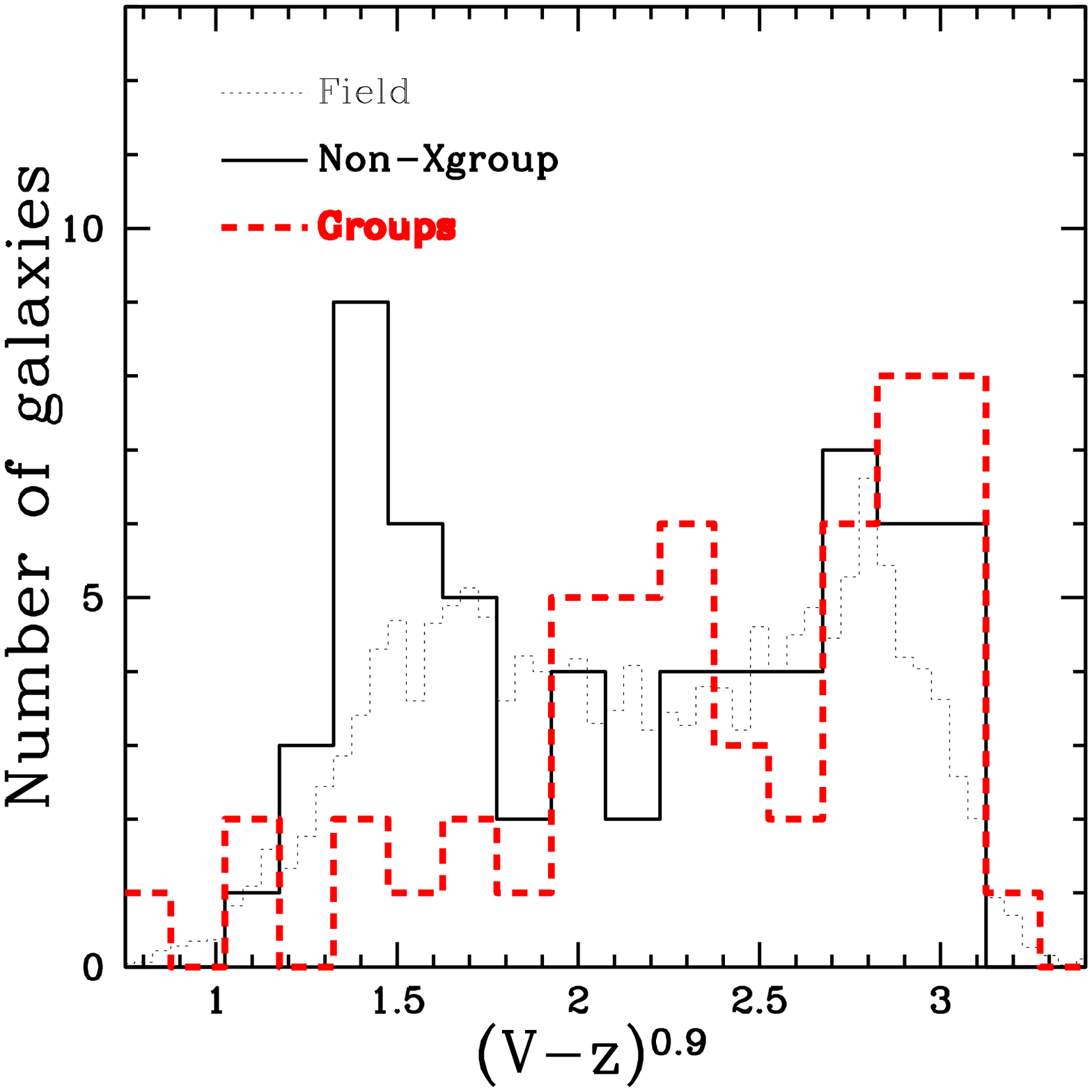} \leavevmode \epsfysize=8cm \epsfbox{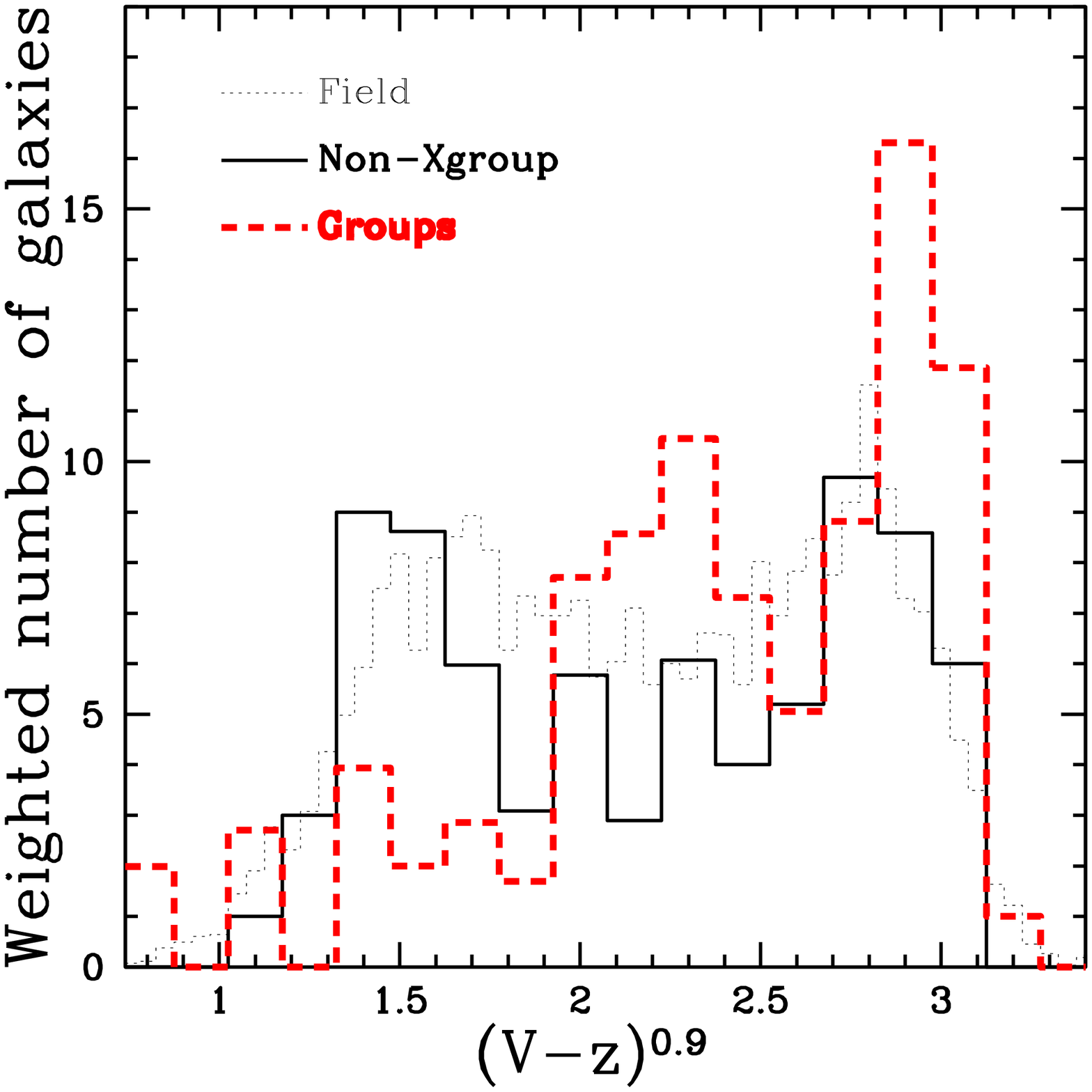}
\caption{{\it Left: }The distribution of $(V-z)^{0.9}$ colours are
  shown, for all galaxies with $M_{\rm star}>10^{10.1}M_\odot$.   The {\it
    solid, black} line represents the distribution of our
  spectroscopic, non-group galaxy sample, which we suggest represents
  the ``non-Xgroup'' environment.  It shows the well-known bimodal
  distribution, consisting of red and blue galaxy populations.  The
  {\it red, dashed} histogram represents our spectroscopic group
  sample; there are few blue galaxies and, instead the population is
  dominated by red-sequence galaxies, and an apparently distinct
  population of intermediate-colour (``green'') galaxies.  Finally, the 
  {\it thin, dotted black}
  histogram is the distribution of all galaxies in COSMOS, with
  $0.8<z_{\rm phot}<1.0$, $r<24.75$ and $M_{\rm star}>10^{10.1}M_\odot$; this has been renormalized to match the
  area of the red histogram.   This represents the global, ``field''
  population of galaxies.  {\it Right: }The same,
  but where the group sample is now weighted for sampling
  incompleteness, and the COSMOS comparison sample is appropriately
  rescaled.  \label{fig-cdist} }
\end{figure*}
Our main result from these first data is shown in
Figure~\ref{fig-cmass_z}, as
the correlation between $(V-z)^{0.9}$ colour and stellar mass for
the group and non-Xgroup samples.  The small, black points
represent galaxies in the entire COSMOS photometric redshift catalogue,
within $0.8<z_{\rm phot}<1.0$ and restricted to
$r<24.75$, the GEEC selection limit.  In the right panel, 
the large points correspond to
spectroscopically confirmed group members, as defined in
\S~\ref{sec-gmem}.   On the other hand, large points in the left panel
represent all galaxies with secure redshifts $0.8<z_{\rm spec}<1$ that are unassociated with
a group in our catalogue.  We will refer to this spectroscopic sample
as ``non-Xgroup'' galaxies, since they are confirmed non-members of our
X--ray group sample.  We cannot exclude the possibility that
this sample still contains galaxies that are members of other groups,
not identified in X-ray emission.
  The solid line in
Figure~\ref{fig-cmass_z} indicates our colour-dependent completeness
limit, as a consequence of the $r>24.75$ mag selection.  This is
computed by simply looking at the average stellar mass of galaxies near
the magnitude limit, as a function of colour.

We first note the usual structure in this colour-magnitude diagram,
with a ``red sequence'' and ``blue cloud'' population.  We show the
approximate locations of these populations with red and blue ellipses,
respectively.   The red sequence in the groups is well populated at all
masses above our completeness limit, $M>10^{10.1}M_\odot$.  In fact,
there is some indication that the lowest-mass ($M<10^{10.5}M_\odot$), red galaxies appear {\it
  only} in groups, as they are absent in the ``non-Xgroup'' sample.
However the sample is too small to be definitive about this, and we will defer a full analysis of the luminosity function
shapes to the end of the survey. 

The second interesing observation from Figure~\ref{fig-cmass_z} is that
the blue sequence appears to be very underpopulated amongst group
galaxies, compared with the field, for $M\gtrsim10^{10.2}M_\odot$.  The
blue galaxies that do exist here tend to lie at the red edge of the
cloud; moreover, the ``green valley'' between the two sequences is
well-populated, a point we will return to below. 

We have matched our redshift catalogue with the deep {\it
  Spitzer} MIPS
catalogue, which has a 5$\sigma$ detection limit of
$0.071$ mJy \citep{COSMOS_Spitzer}.  Using the calibration of
\citet{Rieke+09}, this limit corresponds
to a star formation rate of SFR$\sim11.8 M_\odot\mbox{yr}^{-1}$ at $z\sim 1$.  These 24$\mu$m
detections are  circled
in red in Figure~\ref{fig-cmass_z}.  The non-Xgroup sample has a large
population of massive, blue galaxies with 24$\mu m$ detections,
indicating substantial star formation rates.   In the groups, the
massive 24$\mu$m detections tend to be considerably redder, populating
the red envelope of the blue cloud, and the green valley itself. \begin{figure*}
\leavevmode \epsfysize=8cm \epsfbox{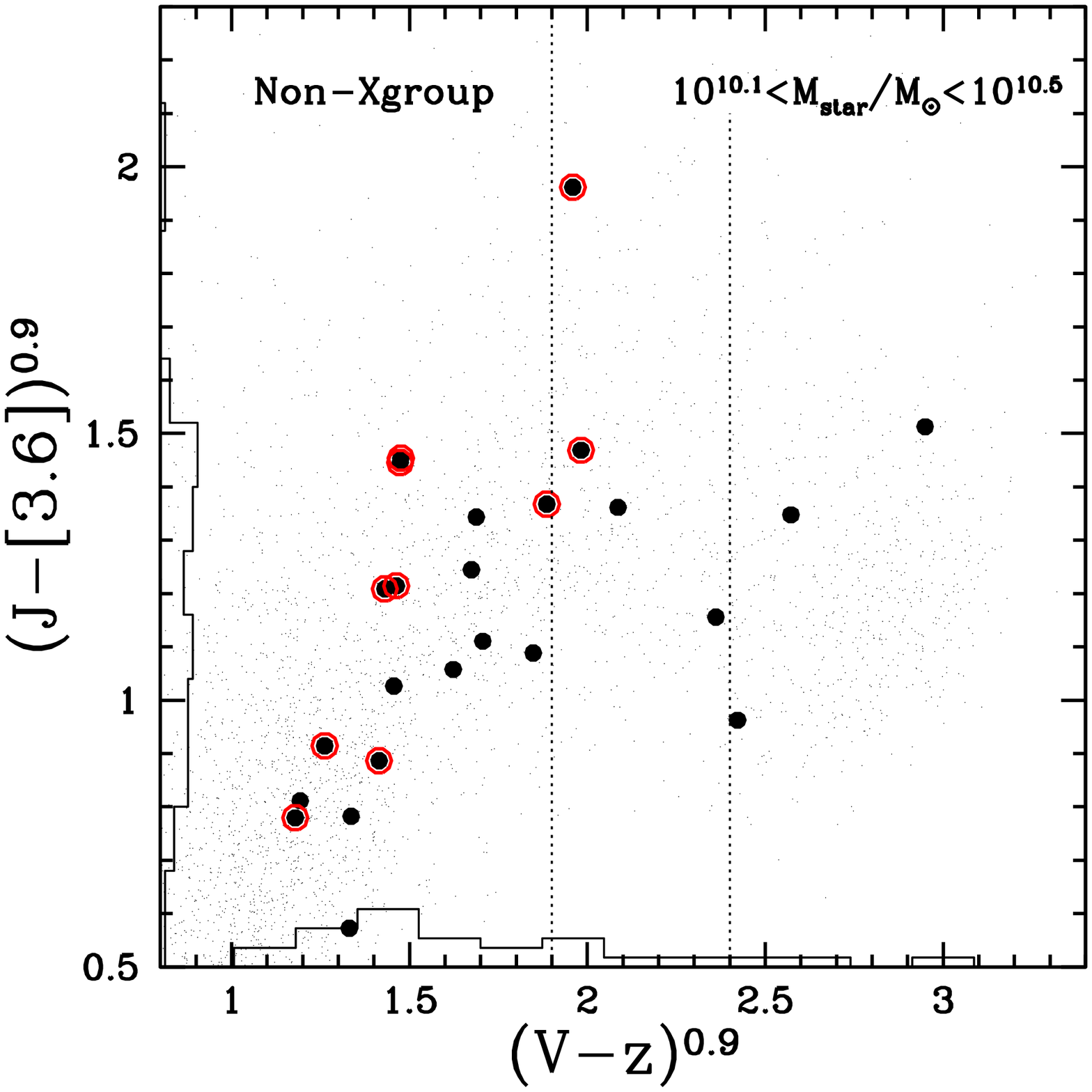}\leavevmode \epsfysize=8cm \epsfbox{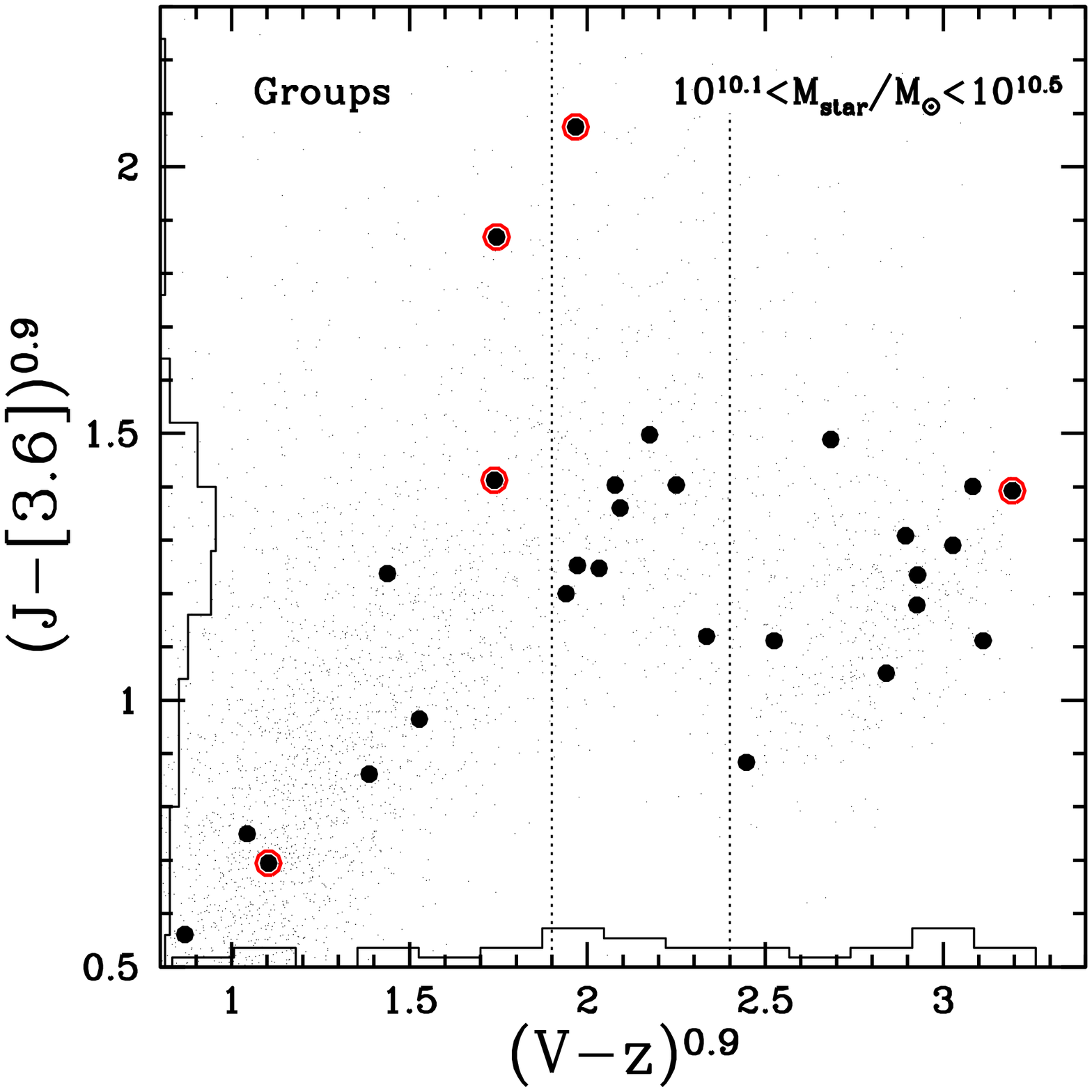}
\caption{Colour-colour plots, aimed at separating dust-reddened spiral
  galaxies from the truly passive population
  \citep[e.g.][]{geec_colours}.  The small, black points represent a
  random third of the
  COSMOS sample (limited to $r<24.75$), while the large points are
  our spectroscopically confirmed galaxies.  Confirmed group members
  are shown in the {\it right} panel, while non-group members (the
  ``non-Xgroup'' spectroscopic sample) are shown on the {\it left}.  In both cases, only
  galaxies with $10^{10.1}<M_{\rm star}/M_\odot<10^{10.5}$  are shown.  Points encircled in red
  represent those group members that are found in the deep MIPS source
  catalogue.  With the exception
  of the single point near the top of the figure, none of the group galaxies in
  the apparently distinct population of ``green'' galaxies (between the
  dashed lines) have the red $(J-[3.6])^{0.9}$ colours expected of
  dusty, star-forming galaxies.  The histograms on either axis just
  represent the arbitrarily normalized distribution of each colour; the
  intermediate population is only apparent in colours that bracket the
  4000\AA\ break.\label{fig-ccbest} }
\end{figure*}



The population of galaxies between the red and blue galaxies is of
particular interest, as it may represent a transient phase.
To explore
this further, we show the distribution of colours, for all galaxies
with $M{\rm star}>10^{10.1}M_\odot$, in Figure~\ref{fig-cdist}.  The black, solid
histogram represents the distribution for our ``non-Xgroup''
spectroscopic sample; while the sample size is small, it shows the well-known bimodal distribution of
colours, with a minimum around  $(V-z)^{0.9}\sim 2.3$.  This is
statistically consistent\footnote{A KS test shows there is a 38 per
cent probability that the two distributions are drawn from the same
parent population.} with the colour distribution of the full
COSMOS ``field'' sample (for $0.8<z_{\rm phot}<1.0$) in the same
stellar mass range, shown as the thin, dotted line.  The
red, dotted histogram represents our spectroscopic group sample.   The
left panel shows these data unweighted for completeness, while the
right panel includes the completeness corrections.  In both cases, the
groups show a marked deficit of blue galaxies, while the
intermediate-colour population is at least as abundant as in the general
field.  

Remarkably, the distribution suggests that this intermediate
population may be distinct from both the red peak and blue cloud.
Unfortunately the sample is not yet quite
large enough to unambiguously identify these ``green'' galaxies with the group
environment; a KS
test on the unweighted distributions in Figure~\ref{fig-cmass_z}
confirms that the group distribution has a $\sim 1$ per cent
probability of being drawn from the same distribution as the parent,
field population; or a $\sim 2.5$ per cent chance of being drawn from
the spectroscopic, ``non-Xgroup'' sample.    
It is also interesting that the ``green'' population (with $1.9<(V-z)^{0.9}<2.4$) makes
up a similar fraction of the total in the general field ($\sim 25$ per
cent) as in the groups ($\sim 30$ per cent).  The main difference
between the two distributions is that, in the
groups, the bluer peak contains only $\sim 10$ per cent of the
galaxies, while the fraction is $\sim 35$ per cent in the field.   

Whether or not it is unique to galaxy groups, the existence of this population of
galaxies with intermediate colours is likely robust.   Galaxies of this
colour are found  in all seven of our groups, and
we have checked
that the distribution in Figure~\ref{fig-cdist} is not greatly affected
by the k-corrections, redshift quality cut or group membership
determination parameters.
In
Figure~\ref{fig-ccbest} we show colour-colour diagrams of our
spectroscopic samples, again compared with the field sample from
COSMOS.  We show both the ``non-Xgroup'' spectroscopic sample, and the
confirmed group members, in separate panels, and we restrict the plots to galaxies with
$10^{10.1}<M_{\rm star}/M_\odot<10^{10.5}$ to focus on the lowest-mass galaxies
for which our sample is complete.  The second colour here, $(J-[3.6])^{0.9}$, is chosen to sample
approximately rest-frame $(R-H)$, a colour that should be sensitive to
dust \citep[e.g.][]{C17-dust,geec_colours}.  Indeed, in the parent
sample we can clearly see the star-forming (blue) sequence extends to
red colours in $(V-z)^{0.9}$; but they remain distinct from the passive
population by virtue of their very red $(J-[3.6])^{0.9}$ colours.

\begin{figure*}
\leavevmode\psfig{figure=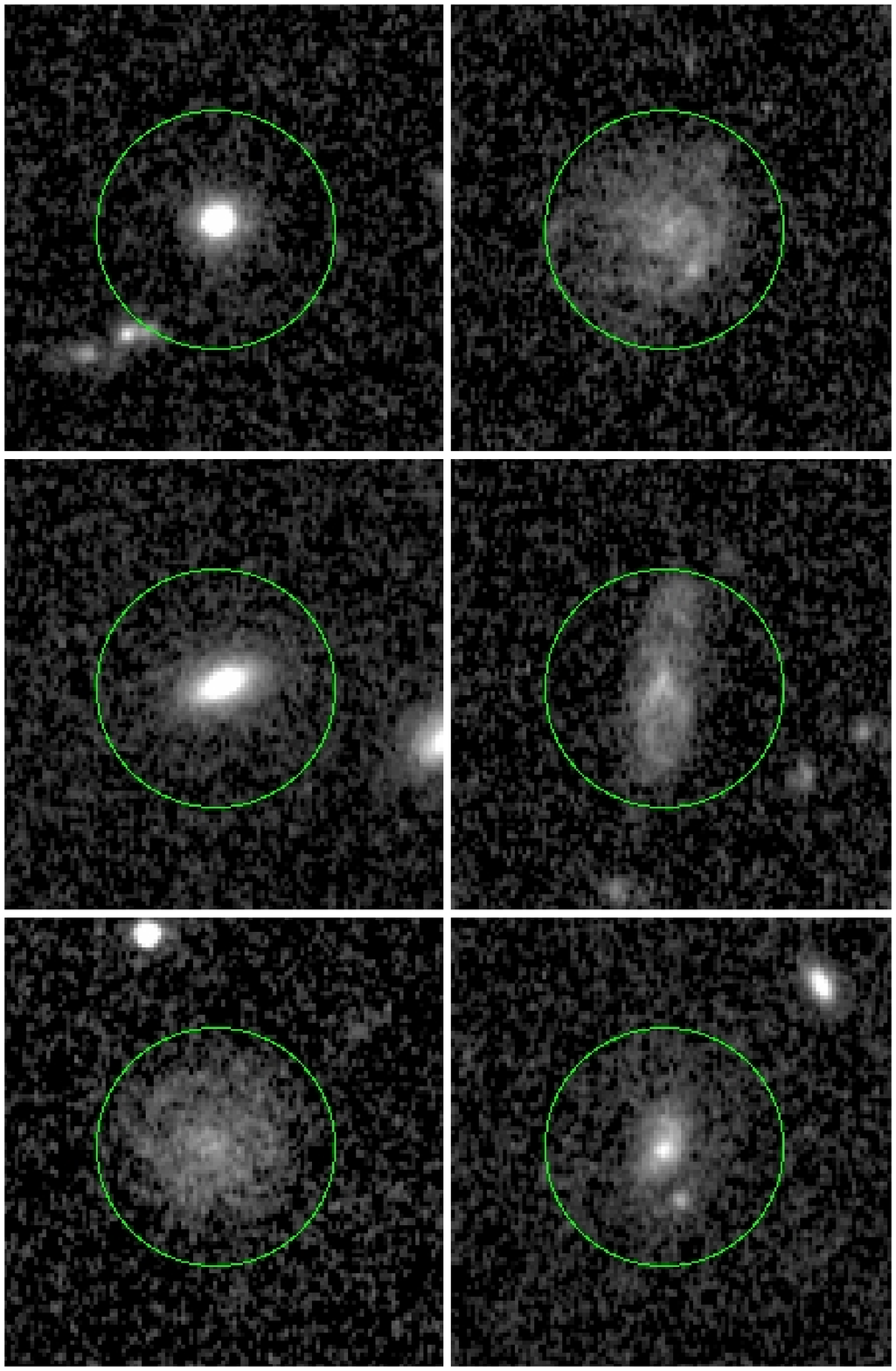,height=6.25cm,width=4.27cm}\hskip .25cm \psfig{figure=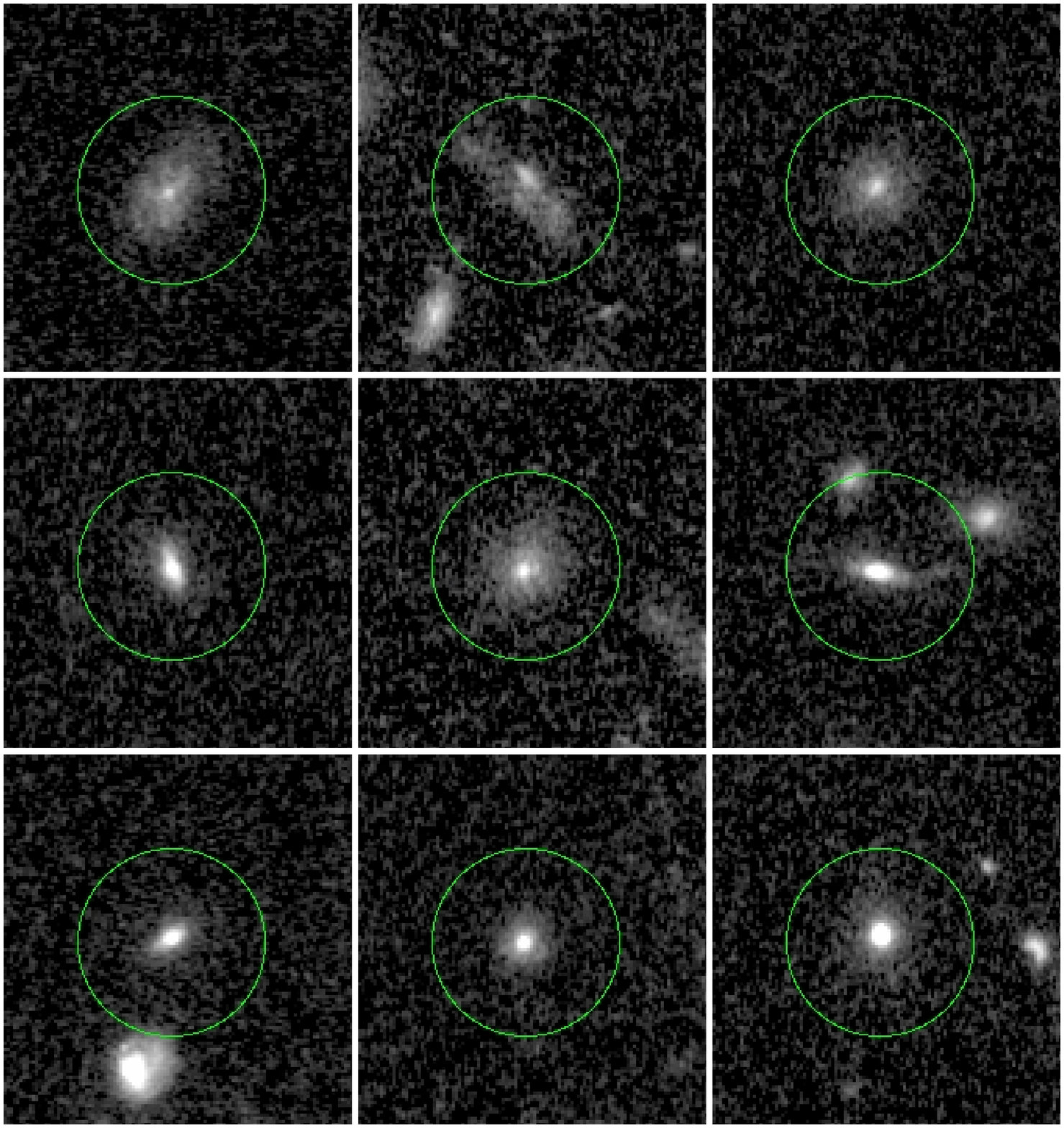,height=6.25cm,width=6.25cm}\hskip .25cm\psfig{figure=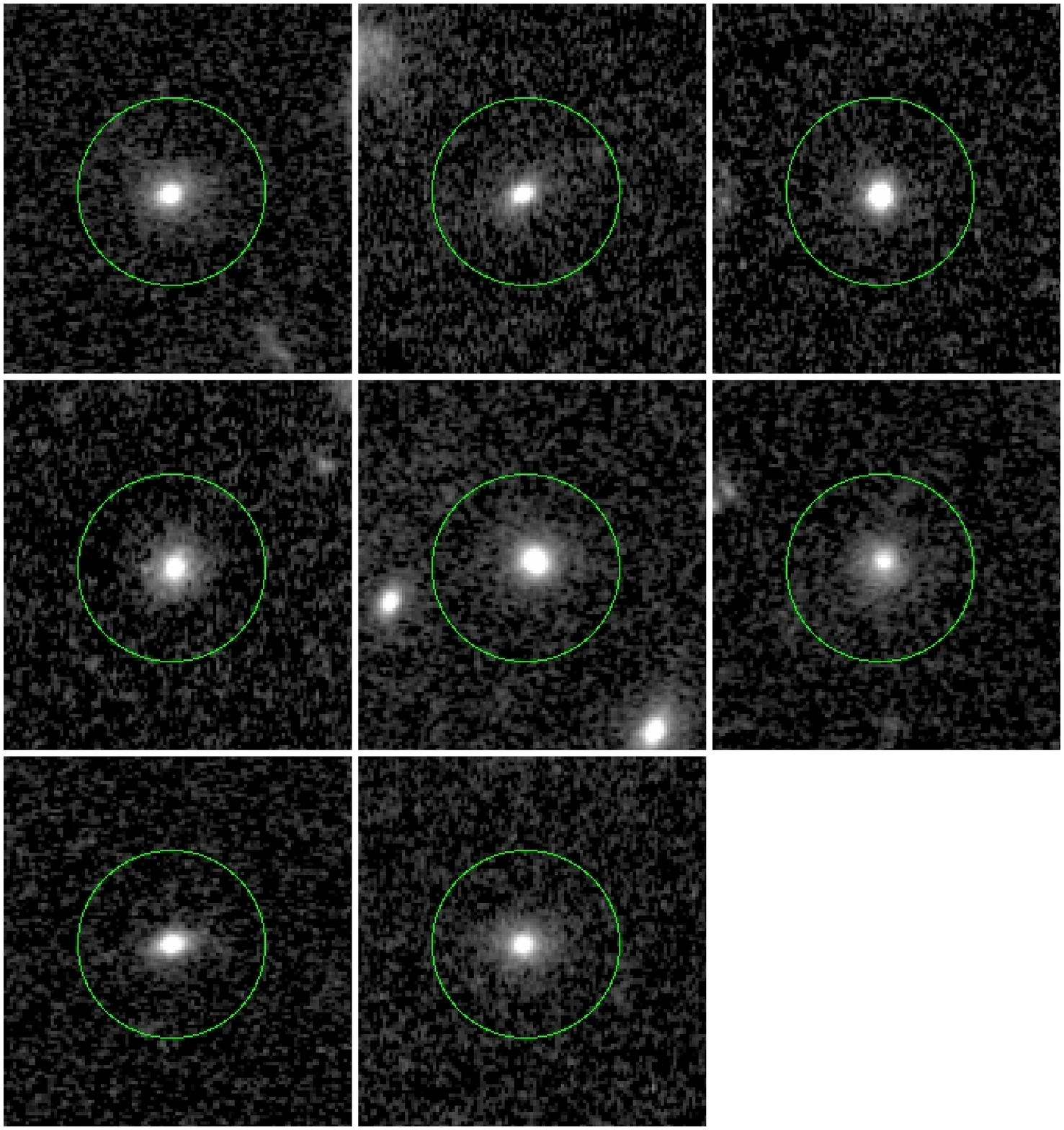,height=6.25cm,width=6.25cm}
\caption{The {\it HST} F814W images of spectroscopically confirmed group
  galaxies in our sample, with $10.1<\log{M_{\rm star}/M_\odot}<10.5$.  The {\it
    left}, {\it middle} and {\it right} panels show blue, green and
red galaxies respectively, as defined in the text.  All images are 6.0
arcsec on each side (47 kpc at $z=0.9$), and the scales and colourbars
are matched, with a square-root scaling applied.  Colours are measured
within a 3\arcsec\ diameter aperture, indicated by the green circles.
In each panel,
galaxies are sorted by colour; galaxies get progressively redder moving
left-to-right, top-to-bottom.
The
green galaxy population appears morphologically distinct from both the
blue and red population.  Most of them possess disks, though they are
smaller and less structured than those of the blue population.  The
exception is the edge-on disky galaxy (top row, middle column), that is
the MIPS--detected galaxy lying on the  dusty star-forming sequence of
Figure~\ref{fig-ccbest}.  \label{fig-morphs}}
\end{figure*}

Interestingly, most of the ``green'' galaxies have relatively blue
$(J-[3.6])^{0.9}$ colours, suggesting that they are not exceptionally
dusty. The only exception in the group population is the single
galaxy near the top of the figure, which is in fact an edge-on 
spiral galaxy (see Fig~\ref{fig-morphs} and later discussion).  Interestingly, the distinct
population of ``intermediate'' galaxies is not apparent in the
$(J-[3.6])^{0.9}$ distribution alone.  We have checked using other
filter combinations, and confirm that it is only apparent in colours
that bracket the 4000\AA\ break.

The matches to the deep MIPS catalogue are shown as red
circles in Figure~\ref{fig-ccbest}.   Few of our group galaxies are
detected; in particular of the nine ``green'' galaxies in the mass
range shown here, only one is detected at 24$\mu$m.
This is the edge-on, dusty spiral with very red
$(J-[3.6])$ colour noted previously.  It has a 24$\mu$m flux of $\sim
0.24$ mJy which, using the calibration of \citet{Rieke+09}, corresponds
to a SFR$\sim 78 M_\odot\mbox{yr}^{-1}$. 
\begin{figure}
\leavevmode \epsfysize=8cm \epsfbox{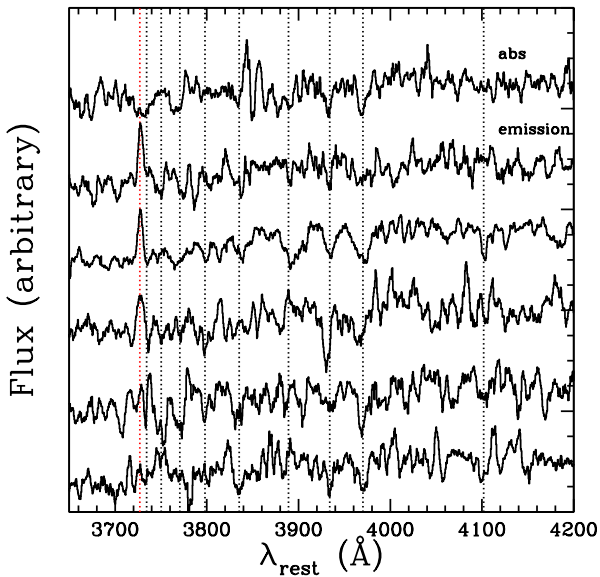} 
\caption{Selected spectra of ``green'', spectroscopically confirmed group
  members.  The spectra have been deredshifted, and smoothed.  Vertical
  lines show the rest-frame wavelengths of common absorption lines (the
  Balmer series and Ca H\&K), and
  of [OII] emission (3727\AA, in red).  The four galaxies on the bottom
  have prominent high-order Balmer absorption lines, possibly
  indicative of recently truncated star formation.\label{fig-plotselect}}
\end{figure}

With the full
sample in hand, we will be able to combine spectra for galaxies with
similar stellar masses and colours, to do a careful line strength
analysis.  For now, we have qualitatively checked the individual spectra of the 
confirmed group members which lie in this intermediate--colour
region.   About half of the galaxies show [OII] emission, although it is weak
in all cases.  Surprisingly, many of the galaxies have prominent H$\delta$
absorption lines, and higher-order Balmer lines are also often apparent.  
This is intriguing and suggests these galaxies may be associated with the
recent cessation of star formation \citep[see
also][]{Ediscs_psb,DEEP2_E+A}.   In particular,
\citet{Ediscs_psb2} find that galaxies with strong H$\delta$ absorption
may be especially numerous
in those groups in the EDisCS sample that have a low fraction of
star-forming galaxies.  
We show six of the better-quality
spectra, in Figure~\ref{fig-plotselect}.  The spectra have been
deredshifted, and boxcar-smoothed over 11 pixels, corresponding to a
rest-frame dispersion of 5.2\AA\ per smoothed pixel.  The bottom two
spectra have strong absorption lines, including H$\delta$, and no significant [OII]
emission.  The next two show both emission and strong absorption lines,
while the top two show an ordinary emission-line and absorption line
galaxy.  We will defer a quantitative analysis of these line indices to
completion of the sample, where we can increase the signal-to-noise
ratio by combining spectra for different classes of galaxies \citep{Dressler04}.

Finally, we consider the morphologies of the spectroscopic group
members, using the {\it HST} image cutouts shown in Figure~\ref{fig-morphs}.  We just consider galaxies
with $10.1<\log{M_{\rm star}/M_\odot}<10.5$ and secure redshifts, and divide them
into blue ($1.2<(V-z)^{0.9}<1.9$), green  ($1.9<(V-z)^{0.9}<2.5$) and
red  ($2.7<(V-z)^{0.9}<3.2$) galaxies.  We defer a detailed discussion,
including surface-brightness profile fitting, to later papers based on
the full sample.  The images here are interesting even qualitatively,
as the three populations appear to be morphologically distinct.  The
small blue population is fairly heterogeneous but includes three galaxies with large, structured
disks, while the
red galaxies are mostly simple elliptical galaxies with little disk
component.  The green galaxies are intermediate --- in general, they are small,
centrally concentrated, smooth and regular in appearance.  Most appear to
have disks, though none are as large or as structured as those found in the blue
population.   This tendency for intermediate-colour galaxies to 
have intermediate morphologies is also seen in the EDisCS sample at
somewhat lower redshift \citet{Pogg+08}.
The exception is an edge-on disky system, which is the
MIPS--detected source, lying on the dusty-end of the blue-galaxy
sequence in Figure~\ref{fig-ccbest}.  Excluding this galaxy, only
one of the eight shows obvious signs of
interaction with neighbours. 

Thus, although this preliminary sample is small, there is evidence for a
distinct galaxy population, with properties intermediate between the
blue cloud and red sequence.  It is
prominent in our sample of groups and is in fact dominant over the
blue cloud in the
mass range $10.1<\log{M_{\rm star}/M_\odot}<10.5$.

\section{Discussion and Conclusions}\label{sec-discuss}
We have presented first results from our ongoing Gemini spectroscopic survey of galaxy
groups at $z\sim 1$.  Using GMOS in nod \& shuffle mode, with two hour
exposures, we are able to obtain secure redshifts for galaxies up to
1.5 mag fainter than zCOSMOS, and thus obtain $\gtrsim10$ members per group
with stellar masses $M_{\rm star}>10^{9}M_\odot$.  Our final sample will consist of $\sim
20$ groups.  Here we
present the survey strategy, and first results on a sample of seven
groups. 

Our main discovery is that the groups host a population of ``green'' galaxies
with colours and morphology that are intermediate between the usual
blue cloud and red sequence.  While a small fraction of these are
dusty, edge-on spiral galaxies, most appear to be a truly intermediate
population. Interestingly, most of the group population ($\sim 90$ per
cent) is made up of
these  green- and red-sequence galaxies; compared with the general
field the groups are largely devoid of massive, blue galaxies.

The simplest explanation for the presence of these green galaxies is that
they are transients.  Broadly speaking there are two likely scenarios
that might give rise to such a population.  In the first hypothesis,
they represent a normal phase of evolution in most, if not all,
galaxies.  If galaxies form stars episodically, due to variations in
cooling, accretion, and feedback processes, we might expect them to
cycle between the blue and green (or red) population.   In this case,
the main characteristic of the group population is the lack of blue
galaxies, which would suggest that it is the {\it
  rejuvenation} of star formation that has been halted.  This is 
consistent with ``strangulation'' models, in which the normal star
formation in disks of satellite galaxies is undisturbed, but the supply
of fresh gas has been removed
\citep[e.g.][]{LTC,infall}.

Recently there have been indications that AGN inhabit the ``green
valley'' of the  galaxy colour distribution
\citep[e.g.][]{M+07,DEEPII_green,C+10}, although this is somewhat
controversial \citep{X+10}.  This
is tempting evidence that the onset of AGN feedback leads to the
cessation of star formation, as predicted by models
\citep[e.g.][]{Croton05,Bower06}, and in support of our first hypothesis.  Only one of our green, group members
is detected as an X--ray point source; it is one of the more massive
such galaxies, with $M_{\rm star}=10^{10.66}M_\odot$.   Several of the more
massive, green galaxies, with  $M_{\rm star}>10^{10.6}M_\odot$, are detected at
24$\mu$m, and thus may be candidates for obscured AGN
\citep{DEEPII_green}.  However, few of the lower mass candidates are
detected  in the deep MIPS observations, and it is possible that they
have a different origin. 

The second hypothesis is that the dense environment of groups directly
induces a decline in star formation rate (possibly preceded by a burst
of star formation), that causes galaxies to
migrate from the blue cloud to the red sequence.  Recently, several studies of
groups and clusters at this redshift have uncovered evidence of a
galaxy population with specific star formation rates significantly
different from surrounding field galaxies of the same stellar mass.  In
some cases this is seen as a reduced star formation rate
\citep[e.g.][]{V+10}, while in other cases there is an apparent 
 {\it
  enhancement} of star formation
\citep[e.g.][]{Elbaz+07,Muzzin+08,DEEPII_envt_again,I+09,ROLES_tornado,K+10}.
Which effect dominates appears to depend both on stellar mass and local
density \citep{Sobral}, but in any case this provides support
for our second hypothesis.  In this case, we
would expect at least a subset of the green population to exist {\it
  only} in such environments and to have a spectral signature that is
distinct from galaxies undergoing a ``normal'' cycle of star formation.   
%
%
Unfortunately, the present sample is not large enough
to conclusively rule out the possibility that the intermediate
``green''  galaxies found in the GEEC groups are equally
abundant in other, lower-density (or higher-density) environments.
This should be possible with the final sample which, importantly, will
include a larger sample of non-group galaxies with similar selection
properties.  

\begin{figure}
\leavevmode \epsfysize=8cm \epsfbox{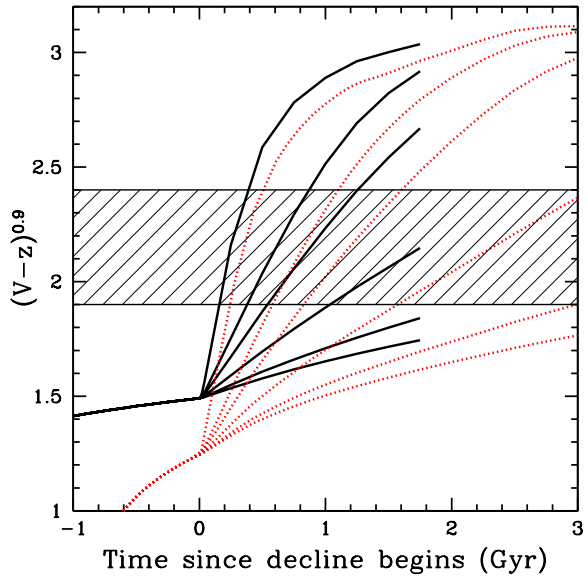} 
\caption{We show predictions of $(V-z)^{0.9}$ colours, from a series of \citet{BC03} stellar
  population synthesis models, assuming a \citet{Chab} IMF and
  [Fe/H]=$-0.33$.  The black, solid lines represent models which have a
  constant star formation rate for 3 Gyr, then exponentially decline
  for another 2 Gyr, with an exponential timescale $\tau$.  The choices
  of $\tau/$Gyr are, increasing from the top curve downward, $0.1$, $0.33$,
  $0.5$, $1$, $2$ and $3$.  The red, dashed lines are similar models,
  but where only the first 1 Gyr are spent with constant SFR, and the
  exponential decline occurs for the
  remaining 4Gyr.   In both cases, the time $t=0$ is defined to be the
  time at which the exponential decline begins.  The shaded region indicates the observed
  colour range of ``green'' galaxies, that lie between the red sequence
  and blue cloud.  Most of the models pass through the ``green'',
  shaded region, and in both cases models with $\tau\sim 1$Gyr spend a
  considerable fraction of their lifetime at this colour.\label{fig-models}}
\end{figure}
\begin{figure}
\leavevmode \epsfysize=8cm \epsfbox{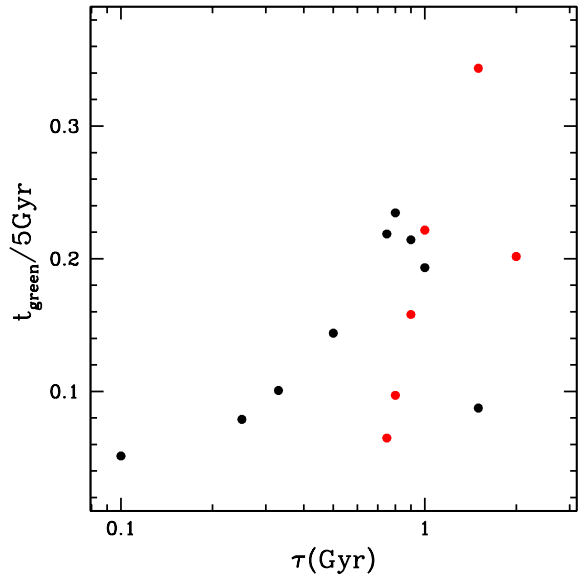} 
\caption{We show predictions from similar models as shown in
  Figure~\ref{fig-models}, with black points representing those that
  begin their decline after 3 Gyr of constant SFR, and red points those
  that decline after only 1 Gyr.  In both cases, the total lifetime is
  5 Gyr, as expected for a galaxy at $z=0.9$ that formed at $z=5$.
  Each point shows the fraction of its lifetime (5 Gyr) that each model
  galaxy spends with ``green'' colours, $1.9<(V-z)^{0.9}<2.4$, as a
  function of $\tau$.  For $0.6<\tau/\mbox{Gyr}<2$, these models spend
  more than 20 per cent of their lifetime as ``green'' galaxies, and
  thus represent a plausible interpretation of our data.
\label{fig-model_green}}
\end{figure}

Assuming that the population of green galaxies represents a one-way
transition between the blue and red populations, we can use their
abundance in groups to estimate the timescale associated with with this
transition.  We have therefore calculated some simple population synthesis models, using the
\citet{BC03} code, with a \citet{Chab} initial mass function and
subsolar metallicity ([Fe/H]=$-0.33$).   We assume a simple extinction model, with
$\tau_v=1$, with 30 per cent of the extinction arising from the
ambient interstellar medium.  
Assuming
galaxies form at $z=5$, the maximum age for a galaxy at $z=0.9$ is
about $5$ Gyr.  For our default models, we assume that the star formation rate is constant for
the first 3 Gyr, after which it begins to decline exponentially with a
timescale $\tau$.    In Figure~\ref{fig-models}, we show the predicted
$(V-z)^{0.9}$ colour as a function of time after this decline begins,
for $\tau/\mbox{Gyr}=0.1, 0.5, 1, 2$ and $3$.  In order to better match the
colours of the red and blue peaks we have applied a $+0.2$ mag offset
to the predicted colours.  
Specifically, after 3 Gyr of
constant star formation all models begin with colours close to the
peak of the observed blue galaxy distribution.  Galaxies with swift
declines ($\tau<1$ Gyr) reach $(V-z)^{0.9}\sim 3.1$, the observed red peak.  

We now turn our attention to the models which are able to reproduce the
observed colours of the ``green'' population, $1.9<(V-z)^{0.9}<2.4$, indicated by the shaded region.  Models with $\tau>1$ Gyr are
effectively ruled out, as the galaxy never gets red enough to reach the
green valley in the time allowed since $z=5$.  On the other hand, models with
$\tau<0.5$ redden rapidly and spend little time in the green valley.
To better quantify this, in Figure~\ref{fig-model_green} we show the
fraction of its 5 Gyr lifetime that these models spend in the green
region, as a function of $\tau$.  All models start in the blue cloud, with 3 Gyr of
constant star formation, which means the maximum possible fraction of
time as a green galaxy is 0.40.  Interestingly, this picks out a
particular timescale, of $0.6<\tau/\mbox{Gyr}<1.3$; such models spend a
significant fraction of time at this colour.  In fact, this fraction
compares reasonably well with the $\sim 30$ per cent of galaxies observed at this colour.

Longer timescales can be accommodated if the truncation starts
earlier.  The red, dashed lines in Figure~\ref{fig-models} show the
results for models in which galaxies form stars at a constant rate for
only $1$Gyr, before beginning their exponential decline.  These
galaxies start off much bluer, at $(V-z)^{0.9}\sim1.3$; otherwise their
behaviour is qualitatively similar to those shown in
Figure~\ref{fig-models}.   The main difference is they have up to 80
per cent of their lifetime available following the truncation.  The red
points in Figure~\ref{fig-model_green} show the fraction of time these
galaxies spend in the green region; in this case, longer timescales of
$\sim 2$ Gyr are preferred, and galaxies can spend up to 35 per cent of
their life at this colour.

These are simplified, parametrized models.  However, the results are
encouraging as they demonstrate it is plausible for a significant
fraction of galaxies to be observed in this special stage of
evolution.   If they really are migrating from the blue to the red
sequence, it is unlikely that they are doing so with star formation
rates declining much more rapidly than $\tau\sim 0.6$ Gyr.  Together
with the undisturbed morphologies, this makes it unlikely that the
transformation is being driven by mergers or rapid gas-depletion
processes.  
Accelerated gas loss through a process like strangulation
\citep[e.g.][]{infall,Ian_rps} predicts a star formation rate decline
on a timescale consistent with our observational constraints.   This
may be evidence that low--mass satellite galaxies have a fundamentally
different evolution in groups, compared with their more massive counterparts, for
which there is some evidence for merger-induced enhancements in star
formation \citep[e.g.][]{K+10}.

One of the next
steps will be to couple the simple models presented here with an
assembly history prediction \citep{McGee-accretion} to achieve better
constraints.  Finally, with a larger sample of both group and isolated galaxies, and a
comparison with more massive galaxy clusters at this redshift, we will
be better able to determine to what extent these transforming galaxies are associated with
the group environment.  


\section{Acknowledgments}\label{sec-akn}
We thank the referee for a helpful report which led to the addition of
several important clarifications and tests.
We are also grateful to the COSMOS and zCOSMOS teams for making their excellent data
products publicly available.  
This research is supported by NSERC Discovery grants to MLB and LCP.  We
thank the DEEP2 team, and Renbin Yan in particular for providing the
{\sc zspec} software, and David Gilbank for helping us adapt this to
our GMOS data.  
\bibliography{ms}
\end{document}